\documentclass[12pt,preprint]{aastex}
\usepackage{xspace}
\usepackage{graphicx}
\usepackage{natbib}
\newcommand{\nraoblurb}{The National Radio Astronomy Observatory is
a facility of the National Science Foundation operated under cooperative
agreement by Associated Universities, Inc.}

\newcommand{\gl}{\ensuremath{\ell}\xspace}
\newcommand{\gb}{\ensuremath{{\it b}}\xspace}

\newcommand{\lb}{\ensuremath{(\gl,\gb)}\xspace}
\newcommand{\lv}{\ensuremath{(\gl,v)}\xspace}

\newcommand{\kms}{\ensuremath{\,{\rm km\,s^{-1}}}\xspace}

\newcommand{\kpc}{\ensuremath{\,{\rm kpc}}\xspace}

\newcommand{\degree}{\ensuremath{\,^\circ}\xspace}

\newcommand{\hi}{{\rm H\,{\footnotesize I}}\xspace}
\newcommand{\hii}{{\rm H\,{\footnotesize II}}\xspace}

\newcommand{\cor}{\ensuremath{^{\rm 13}{\rm CO}}\xspace}

\newcommand{\hiea}{{\rm H\,}{{\footnotesize I}{\rm~E/A}}}

\shorttitle{HRDS Distances}
\shortauthors{Anderson et al.}

\begin{document}

\title{The Green Bank Telescope H{\footnotesize II} Region Discovery Survey:\\ III. Kinematic Distances}

\author{L. D. Anderson\altaffilmark{1},\email{Loren.Anderson@mail.wvu.edu} T. M. Bania\altaffilmark{2}, 
Dana S. Balser\altaffilmark{3}, \& Robert T. Rood\altaffilmark{4}}

\altaffiltext{1}{Department of Physics, West Virginia University, Morgantown, WV 26506, USA.}
\altaffiltext{2}{Institute for Astrophysical Research, Department of Astronomy, Boston University, 725 Commonwealth Ave.,
Boston MA 02215, USA.}
\altaffiltext{3}{National Radio Astronomy Observatory, 520 Edgemont Road,
Charlottesville VA, 22903-2475, USA.}
\altaffiltext{4}{Astronomy Department, University of Virginia, P.O. Box 3818,
Charlottesville VA 22903-0818, USA.}

\begin{abstract}
\noindent
Using the \hi\ Emission/Absorption method, we resolve the kinematic
distance ambiguity and derive distances for 149 of 182 (82\%)
\hii\ regions discovered by the Green Bank Telescope \hii\ Region
Discovery Survey (GBT HRDS).  The HRDS is an X-band (9\,GHz, 3\,cm)
GBT survey of 448 previously unknown \hii\ regions in radio
recombination line and radio continuum emission.  Here we focus on
HRDS sources from $67\degr \ge \ell \ge 18\degr$, where kinematic
distances are more reliable.  The 25 HRDS sources in this zone that
have negative recombination line velocities are unambiguously beyond
the orbit of the Sun, up to 20\,kpc distant.  They are the most
distant \hii\ regions yet discovered.  We find that 61\% of HRDS
sources are located at the far distance, 31\% at the tangent point
distance, and only 7\% at the near distance.  ``Bubble'' \hii\ regions
are not preferentially at the near distance (as was assumed
previously) but average 10\,kpc from the Sun.  The HRDS nebulae, when
combined with a large sample of \hii\ regions with previously known
distances, show evidence of spiral structure in two circular arc
segments of mean Galactocentric radii of 4.25 and 6.0\,kpc.  We perform
a thorough uncertainty analysis to analyze the effect of using
different rotation curves, streaming motions, and a change to the
Solar circular rotation speed.  The median distance
uncertainty for our sample of \hii\ regions is only 0.5\,kpc, or 5\%.  This
is significantly less than the median difference between the near and
far kinematic distances, 6\,kpc.  The basic Galactic structure results
are unchanged after considering these sources of uncertainty.
\end{abstract}

\section{Introduction}


\hii\ regions, the zones of ionized gas surrounding massive OB stars,
have been instrumental to our understanding of the star formation
history, structure, and composition of our Milky Way Galaxy.  While
there are many extant catalogs of \hii\ regions, distance information
is frequently lacking.  Accurate distances are required to turn the
measured properties of flux and angular size into the physical
properties of luminosity and physical size.  Because OB stars have
very brief lifetimes, \hii\ regions trace star formation at the
present epoch.  They therefore are found only in locations of active
star formation, primarily in spiral arms.  Their chemical composition
is that of the present-day interstellar medium, after billions of
years of stellar processing.  Distances are required if we are to use
\hii\ regions to trace Galactic structure or to learn about the
chemical evolution of our Galaxy.

Measured radial velocities can be used to compute kinematic distances
using a rotation curve model for the Galaxy.  Rotation curves usually
assume circular rotation about the Galactic center such that a model
radial velocity is a function only of its Galactocentric distance.
Galactic rotation curves have in general been derived using either CO
\citep[e.g.,][]{clemens85} or \hi\ \citep[e.g.,][]{burtongordon78}.
The different tracers employed and the different methodologies used to
derive the rotation curves from measured velocity fields cause
slightly different results.

Spectro-photometric distances \citep[e.g.,][]{russeil07} and
trigonometric parallax of associated masers \citep[e.g.,][]{reid09}
are potentially more accurate methods for calculating Galactic
\hii\ region distances compared to kinematic distances.  Distances
derived using maser parallax measurements typically have low
uncertainties compared to kinematic distances.  \citet{reid09} quote
an average uncertainty of 10\% for distances of 10\,kpc and found for
some sources discrepancies of over a factor of 2 between the kinematic
and the maser parallax distances.  In an extreme case, G9.62+0.20 has
near and far kinematic distances of $\sim0.5$\,kpc and $\sim16$\,kpc,
respectively, and \citet{sanna09} find a maser parallax distance of
5.2\,kpc.  The Galactic location of this source within 10$\degree$ of
the Galactic center direction, however, implies {\it a priori} that
kinematic distances are not reliable.

The Green Bank Telescope \hii\ region discovery survey \citep[GBT
  HRDS;][]{bania10-paperi, anderson11-paperii} discovered 448 Galactic
\hii\ regions by measuring their radio recombination line (RRL)
velocities and radio continuum emission.  The HRDS sources are found
over $67\degree \ge \ell \ge 343\degree$, $|b| \le 1\degr$ and have
doubled the number of previously known \hii\ regions in this zone.
Little is known about many of these regions.  

Only kinematic distances are possible if we are to derive distances to
the majority of the HRDS sample.  One must locate the exciting star(s)
in the optical or near infrared and assign a spectral type to derive a
spectro-photometric distance.  This is in general not possible for
HRDS sources due to extinction as few of the HRDS nebulae are
optically visible.  Maser parallax distances rely on measurements
using Very Long Baseline Interferometry (VLBI) of bright maser sources
associated with massive star forming regions.  Such maser spots are
not uncommon, but are not present for all star-forming regions.  Only
about 10\% of HRDS sources are associated with detected maser emission
\citep{anderson11-paperii}.  Our group just led an unsuccessful effort
to find 12\,GHz methanol masers associated with a sample of distant
HRDS targets with the GBT (Anderson et al., 2012, in prep.).

Most HRDS sources lie in the portion of the Galaxy interior to
  the Solar orbit, the ``inner Galaxy.''  Each measured inner Galaxy
  velocity corresponds to two distinct kinematic distances, a near
and a far distance.  This problem is known as the kinematic distance
ambiguity (KDA).  Measured velocities for first-quadrant sources
  in the inner Galaxy increase with distance from the Sun until the
tangent point, which is the location where the radial velocity is at a
maximum along a given line of sight.  Beyond the tangent point, radial
velocities decrease.  The near and far distance are spaced evenly
along the line of sight about the tangent point.  There are two cases
over the Galactic range of the HRDS where there is no KDA: 1) sources
whose velocity is the same as the tangent point velocity and 2)
sources whose velocity places them unambiguously beyond the orbit of
the Sun.  In the first Galactic quadrant, sources beyond the orbit of
the Sun have negative velocities whereas in the fourth Galactic
quadrant the same is true for sources with positive velocities.

There are two common methods one can use to resolve the
KDA for Galactic \hii\ regions.  Both of these methods involve the
detection of a spectral line in absorption from foreground material in
the direction of an \hii\ region. \hii\ regions emit broadband thermal
continuum radiation and an absorption signal may be detected for any
spectral line originating from foreground material with a lower
brightness temperature than that of the \hii\ region.  The most robust
such method uses \hi\ as the absorbing material.  This method is
called the \hi\ emission/absorption (\hiea) method \citep{kuchar94,
  kolpak03, anderson09a, urquhart12} and it relies on the detection of
\hi\ absorption at 21\,cm from the continuum emission of an
\hii\ region.  A similar method uses intervening ${\rm H_2CO}$ clouds
instead of \hi\ to search for an absorption signal \citep{wilson72,
  downes80, araya02, watson03, sewilo04}.  Because there is less
  ${\rm H_2CO}$ compared to \hi, this method will more often resolve
  the KDA incorrectly and is applicable to a smaller number of
  \hii\ regions.  Along a given line of sight \citet{watson03} estimate that on average there is
  one ${\rm H_2CO}$ cloud every 2.9\,kpc, whereas \citet{bania84}
  estimate that there is an \hi\ feature every 0.7\,kpc and
  \citet{radhakrishnan72} estimate one \hi\ ``concentration'' every
  0.3\,kpc.  Thus, the ${\rm H_2CO}$ method is unreliable for
    sources within 2.9\,kpc of the tangent point, and the
    \hiea\ method is unreliable for sources within 0.7\,kpc of the
    tangent point.

\citet[hereafter AB]{anderson09a} used the \hiea\ method to resolve the
KDA for a sample of 291 \hii\ regions from $55\degr \ge \ell \ge 16\degr$,
$|b| \le 1\degr$, which represents all \hii\ regions in this zone
known prior to the HRDS.  Excluding the sources with multiple velocity
components and those with RRL velocities within 10 \kms\ of the
tangent-point velocity, they were able to resolve the KDA for 72\% of
these sources using the \hiea\ method.  They found that for the
angularly small ultra-compact and compact \hii\ regions, their success
rate was nearly $\sim85\%$ whereas for larger low-surface brightness
``diffuse'' regions it was only $\sim30\%$.  This work built on
\citet{kuchar94}, who used the \hiea\ method to provide distances for 70
\hii\ regions.



Here we resolve the KDA for 149 HRDS sources using the \hiea\ method
and data from the VLA Galactic Plane Survey \citep[VGPS;][]{stil06}.
The Galactic structure implications will be discussed in a companion
paper (Bania et al., 2012, in preparation).

\section{Galactic Plane Surveys}
The VGPS is a survey of 21\,cm \hi\ emission that extends from
$67\fdg5 \ge \ell \ge 17\fdg9$ at a spatial resolution of $1\arcmin$
and a spectral resolution of 1.56\,\kms.  The RMS noise in the VGPS is
$\sim 2$\,K per 0.824\,\kms\ channel.  To recover the large-scale
emission, the VGPS fills in the zero-spacing information missed with
the VLA with data from the GBT.  In addition to the spectral line data, the
VGPS provides $1\arcmin$-resolution 21\,cm continuum maps from
spectral channels with no line emission.  These maps are vital for the
\hiea\ process employed here.

The HRDS contains RRL and radio continuum measurements for 448 newly
identified \hii\ regions.  \citet[hereafter Paper~I]{bania10-paperi}
give HRDS first science results and \citet[hereafter
  Paper~II]{anderson11-paperii} provide a catalog of the RRL and continuum
properties of the HRDS nebulae.  Over the extent of the VGPS there are
280 HRDS sources.  Ninety-eight of these, however, have multiple RRL
components along the line of sight.  Without additional information,
one cannot derive a kinematic distance to an HRDS source that has
multiple velocity components.
We exclude from the present analysis HRDS sources with multiple RRL
velocity components.
Our final sample of HRDS sources for the present work consists of 182
objects.

\section{The H\,{\footnotesize I}~E/A Method\label{sec:method}}
The \hiea\ method uses the absorption by foreground \hi\ of the background
broad-band \hii\ region continuum emission, which is also bright at
21\,cm, to resolve the KDA.  \hi\ is ubiquitous in the Galaxy and
emits at all allowed velocities.  If the \hii\ region is at the near
kinematic distance, it will show \hi\ absorption features from
0\,\kms\ to the \hii\ region source velocity.  If the \hii\ region is
at the far kinematic distance, it will show \hi\ absorption features
from 0\,\kms\ to the tangent point velocity.  Therefore, if
\hi\ absorption is detected between the \hii\ region velocity and the
tangent point velocity, the \hii\ region must be at the far kinematic
distance.  If \hi\ absorption it is not detected between the
\hii\ region velocity and the tangent point velocity, this favors the
assignment of the near distance.

The above makes the assumption that every sight line has cool
\hi\ in between the near and the far distance.  Testing this
assumption would require extensive modeling to determine the number of
sight lines for which this assumption may not be satisfied.  
Observed Galactic-scale HI emission 
properties are consistent with mean free path between HI featues 
of 0.7\,kpc, so we may expect
that on average the assumption is valid, and especially for sources
with a large difference between the near and the far distances.

The spectrum in the direction of the \hii\ region, the ``on--source''
spectrum, must be compared with a reference ``off--source''
spectrum to distinguish \hi\ absorption from real fluctuations in
\hi\ intensity.  We may express a ``difference'' spectrum:
\begin{equation}
\Delta T(v) = T_{\rm off}\,(v) - T_{\rm on}\,(v) = T_{\rm off}(v) - (T_{\rm off}(v) + T_c - T_c\, e^{-\tau(v)}) = T_{\rm c}[1 - e^{-\tau(v)}]\,,
\end{equation}
where $T_{\rm on}(v)$ and $T_{\rm off}(v)$ are the on-- and
off--source \hi\ intensity at velocity $v$, $T_{\rm c}$ is the
continuum brightness temperature of the \hii\ region, and $\tau(v)$ is
the optical depth of the absorbing gas at velocity $v$ \citep[see, e.g.,][]{kuchar94}.  This assumes
that, aside from the continuum emission, for each velocity the ``on'' and the ``off''
directions have the same intensity.  In this
formulation, absorption features appear as positive values of $\Delta
T(v)$.  The method for creating the on-- and off--source spectra
employed here and explained below is the same as that used by AB.

We estimate the uncertainty in each $\Delta T(v)$ spectrum to help determine
whether a given absorption signal is a real feature or whether it is
caused by noise.  There are two sources of noise that we
consider: instrumental noise and real small-scale spatial fluctuations in
the \hi emission that can mimic absorption signals.
Following AB, we use a single value of the receiver noise for all spectral channels,
$\sigma_{rms}$.  We calculate $\sigma_{rms}$ as the standard deviation
of all off-source spectral channels devoid of emission.  We estimate the noise from
small-scale fluctuations in the \hi\ emission, $\sigma_{T}$, by computing the standard deviation of
values in the off--source spectrum at each velocity:
\begin{equation}
\sigma_{T} {\rm (}v {\rm )} = \left \{ \frac{1}{N}
\displaystyle\sum_{i=0}^n [T_{{\rm off,} i}(v) - \overline{T}_{\rm off}(v)]^2
\right \}^ {1/2},
\label{eq:errors}
\end{equation}
where the summation is carried out over all spectra in the off--source
region and $\overline{T}_{\rm off}(v)$ is the average value of the
off--source spectra at velocity $v$.  As AB did, we estimate the
total uncertainty at each velocity as the greater of $5\sigma_{\rm
  rms}$ and $\sigma_T(v)$, similar to what has been used by other
authors \citep{payne80, kuchar94}.  To be considered a possible
  absorption signal, as opposed to instrumental noise or a background fluctuation,
we require that any absorption is greater than this total
uncertainty. We verify that all possible absorption features have
  the same morphology as the \hii\ region radio continuum emission
  (see below), and so the true significance of a detection is greater
  than that implied by the error analysis.

\begin{figure}
\centering
\includegraphics[width=3 in]{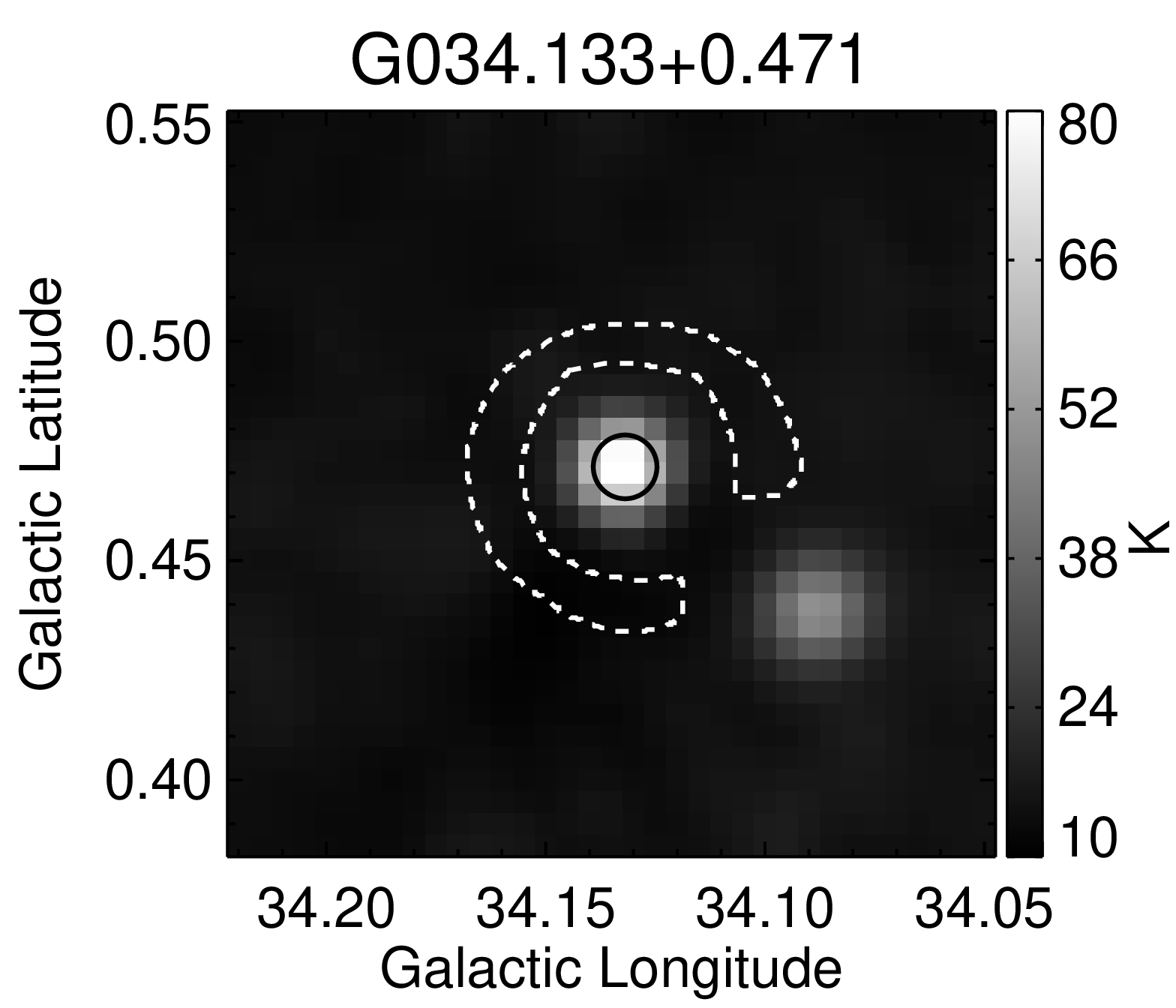}
\includegraphics[width=3 in]{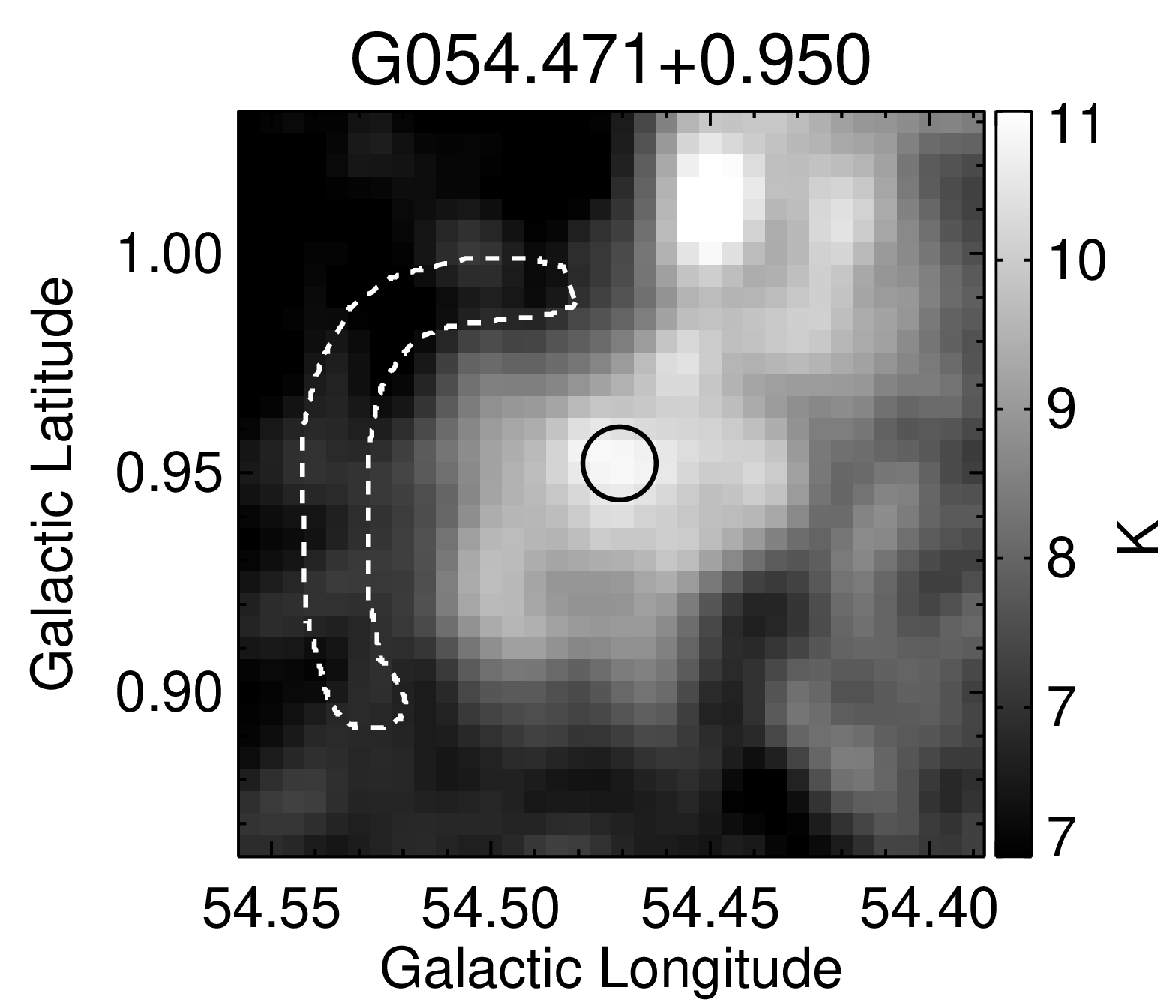}
\includegraphics[width=3 in]{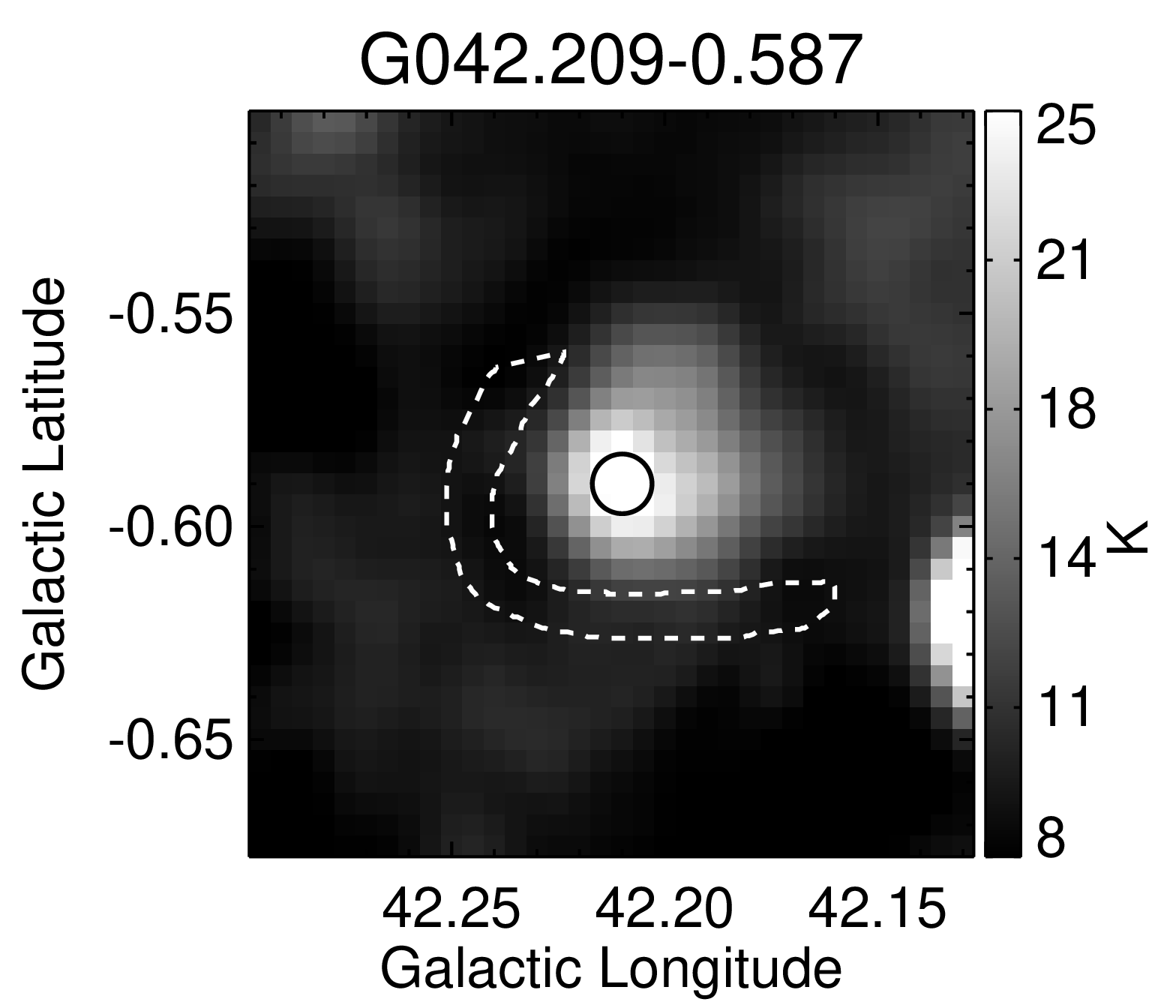}
\includegraphics[width=3 in]{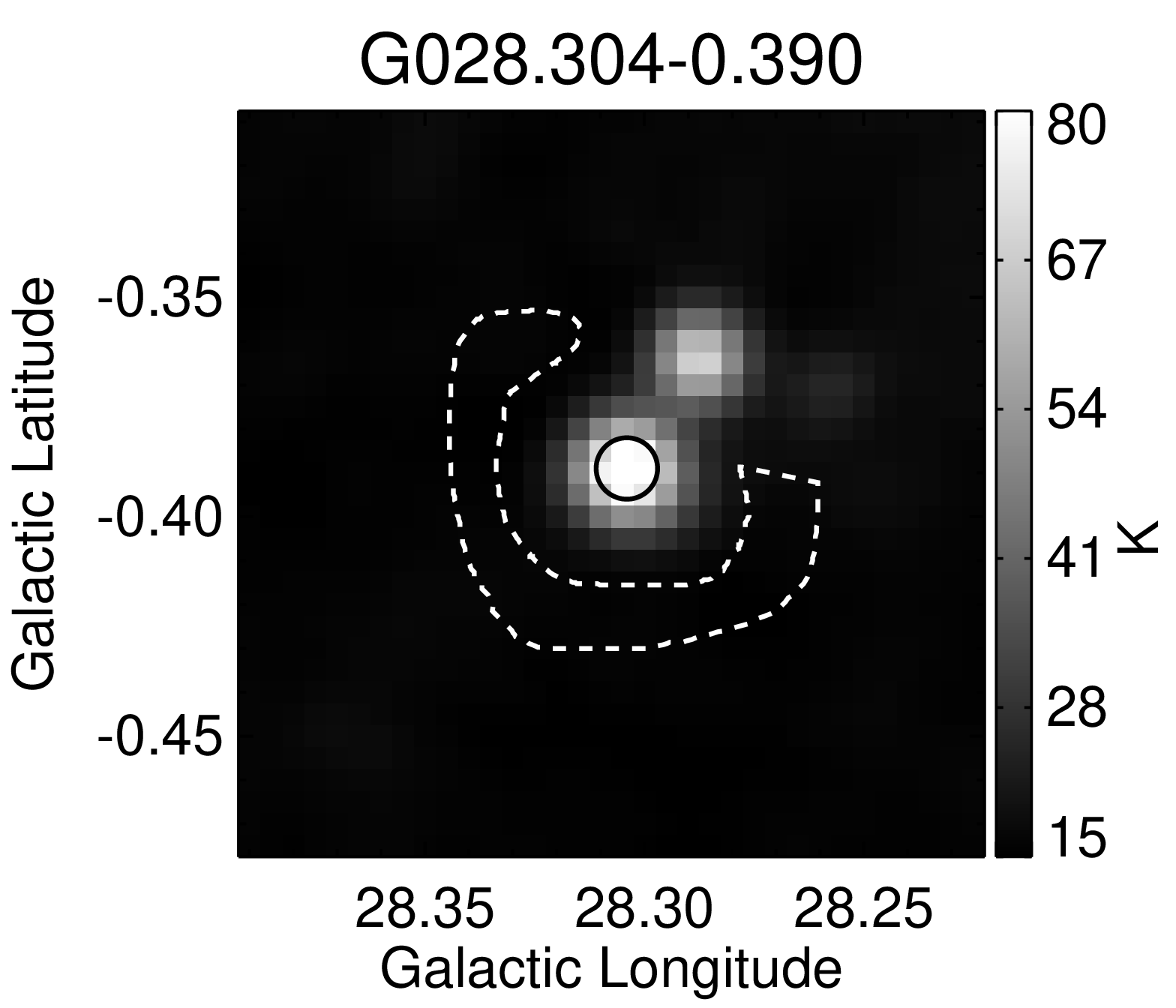}
\caption{On-- and off--source apertures for two \hii\ regions.  The
  on--source apertures are shown with solid lines and the off--source
  apertures with dashed lines.  The background images are VGPS 21\,cm continuum
  data.}

\label{fig:apertures}
\end{figure}

Using the VGPS continuum images as a guide, we define on-- and
off--source apertures with the Kang
software\footnote{http://www.bu.edu/iar/kang/}.  This software allows
the definition of completely arbitrary apertures, which is beneficial
for sources with complex continuum geometries or that are in
complicated regions of emission.  There are two main goals when
defining which ($\ell, b$) areas to use for the on-- and off-- source
regions: the defined apertures should produce spectra with 
the strongest possible absorption signal and the lowest possible
uncertainty due to the combination of instrumental noise and sky 
fluctuations.  To some
extent these goals are contradictory -- the strongest absorption
signal possible will be caused by extracting the spectra from the
single location of brightest radio continuum emission, but this
spectrum will have high instrumental noise.  One can obtain spectra with low
instrumental noise by averaging over a large area, but this will decrease any
absorption signal.  Through repeated trials we found that the best
results were produced with small on-source areas, which maximize the
absorption strength, and larger off--source areas, which minimize the
RMS noise in the off--source spectra.  As AB did, we select the
off--source area such that it surrounds the on-source area but does
not include emission from other discrete radio continuum sources.  To
minimize mis-characterizing real small-scale fluctuations as
absorption, we define the on--source and off--source regions as close
as possible on the sky.  Example on-- and off--source apertures are
shown in Figure~\ref{fig:apertures}.

Using the Kang software we then calculate on-- and off--source spectra
by averaging the spectral line data at the \lb\ pixel locations
falling within the on-- and off--source apertures, respectively.  We
subtract the average on--source spectrum from the average off--source
spectrum to create a difference spectrum, $\Delta T(v)$, that shows
absorption as positive features and compute the uncertainties in the
difference spectra as in Equation~\ref{eq:errors}.

We do not perform a KDA resolution for sources whose velocity is
within 10\,\kms\ of the tangent point velocity, but instead assign these
sources to the tangent point distance.  This affects 36 HRDS sources.
For sources near the tangent point, the distance between the
\hii\ region and the tangent point location is small and thus the
reliability of the \hiea\ method is compromised.  At $\ell=18\degr$,
there is 0.8\,kpc from the tangent point to the distance corresponding
to 10\,\kms\ from the tangent point, according the \citet{brand86}
curve.  Since there is an \hi\ feature on average every 0.7\,kpc along a
given line of sight \citep{bania84}, a KDA resolution using the
\hiea\ method is not reliable for sources within 10\,\kms\ of the
tangent point velocity at $\ell=18\degr$.  At higher longitudes, this
distance increases and the 10\,\kms limit is more conservative.

\begin{figure}
\centering
\includegraphics[width=3 in]{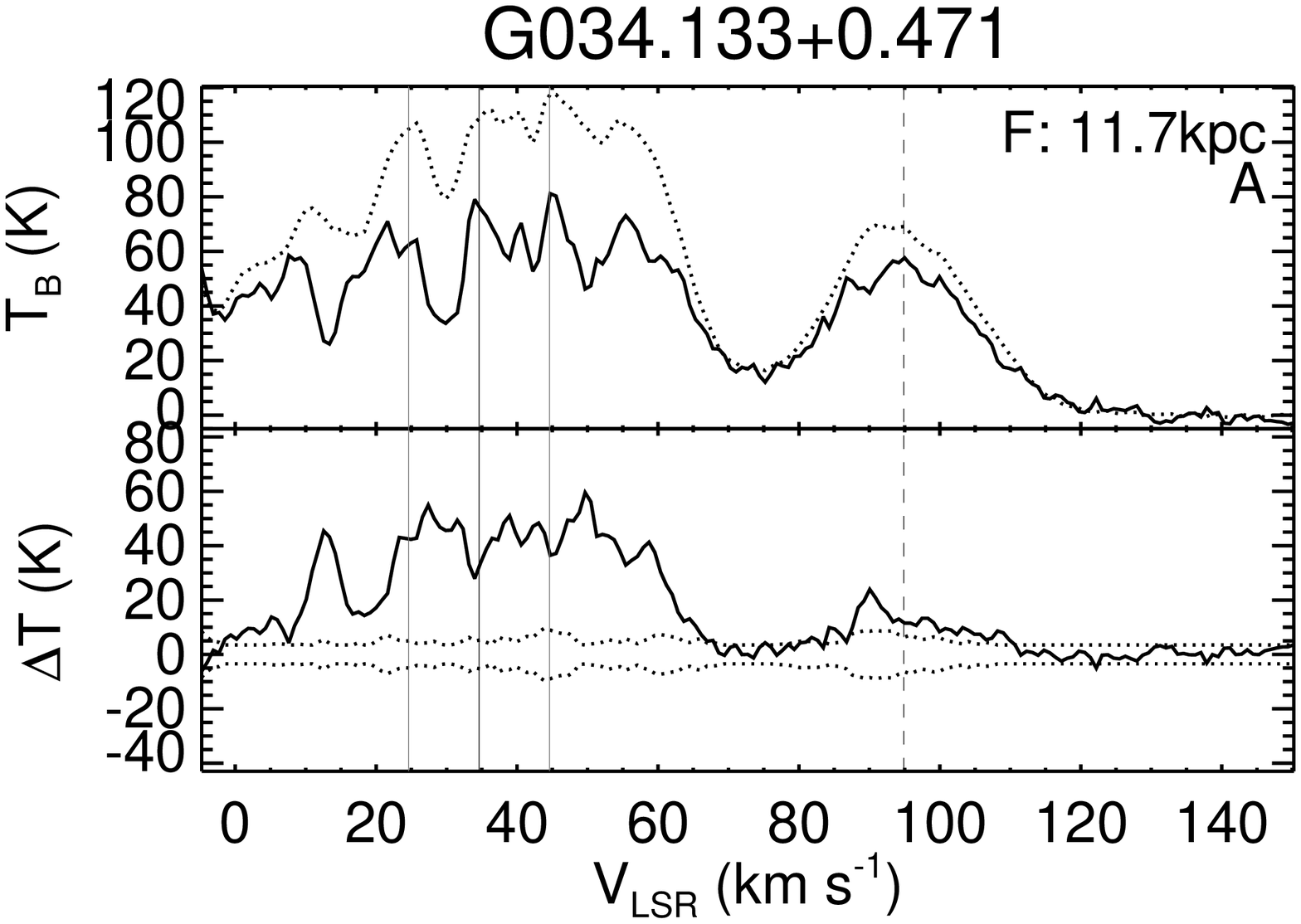} 
\includegraphics[width=3 in]{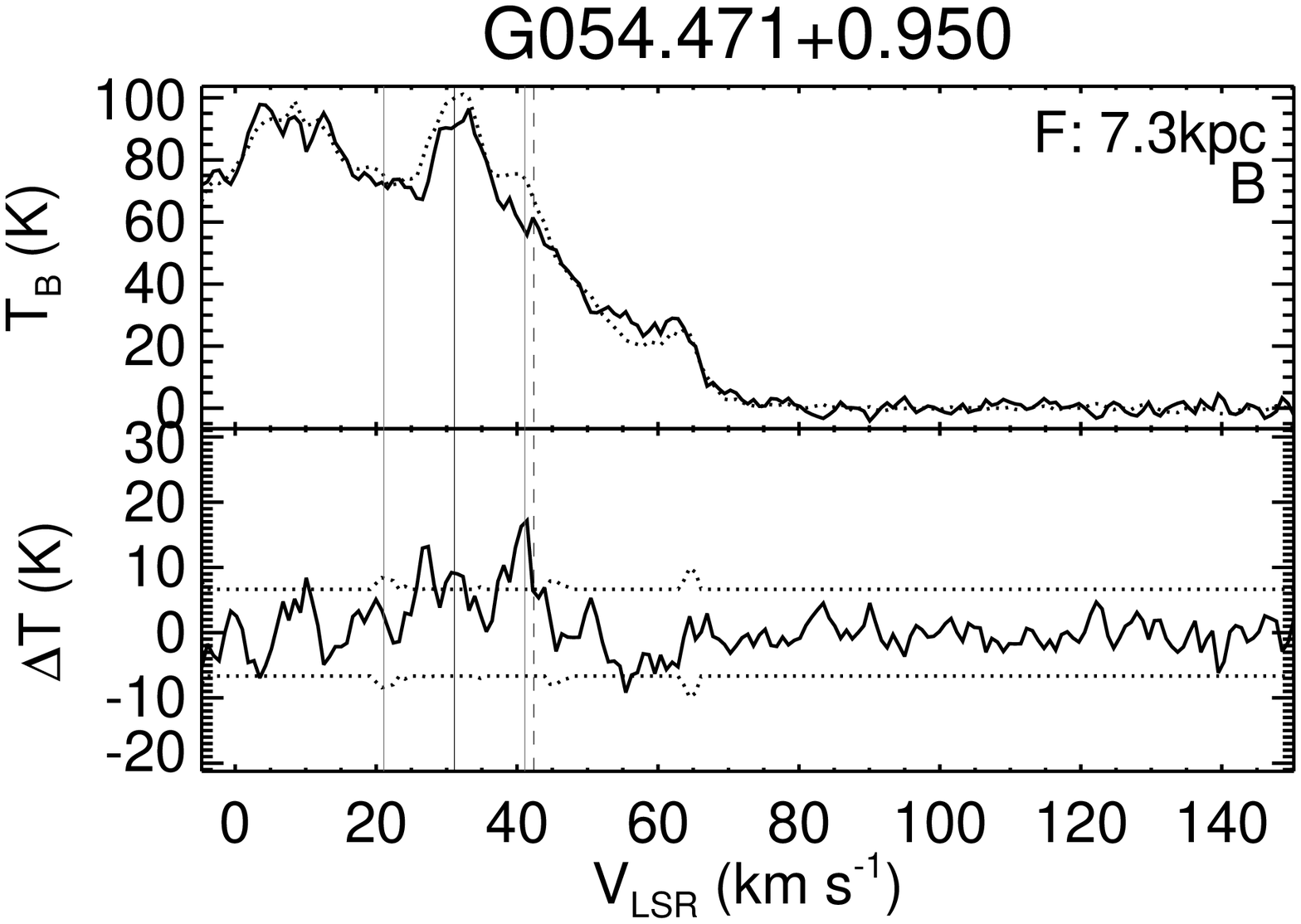} 
\includegraphics[width=3 in]{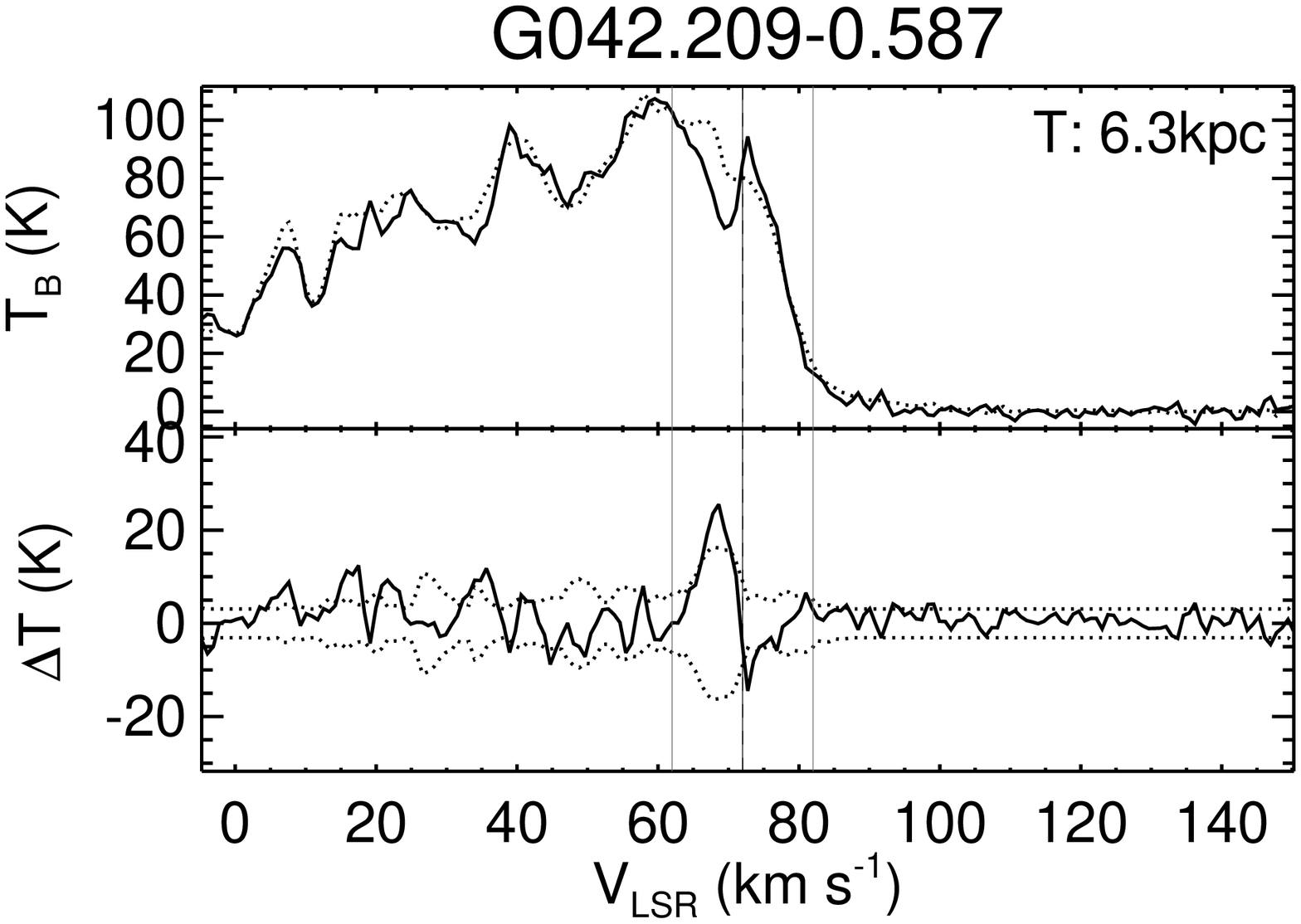} 
\includegraphics[width=3 in]{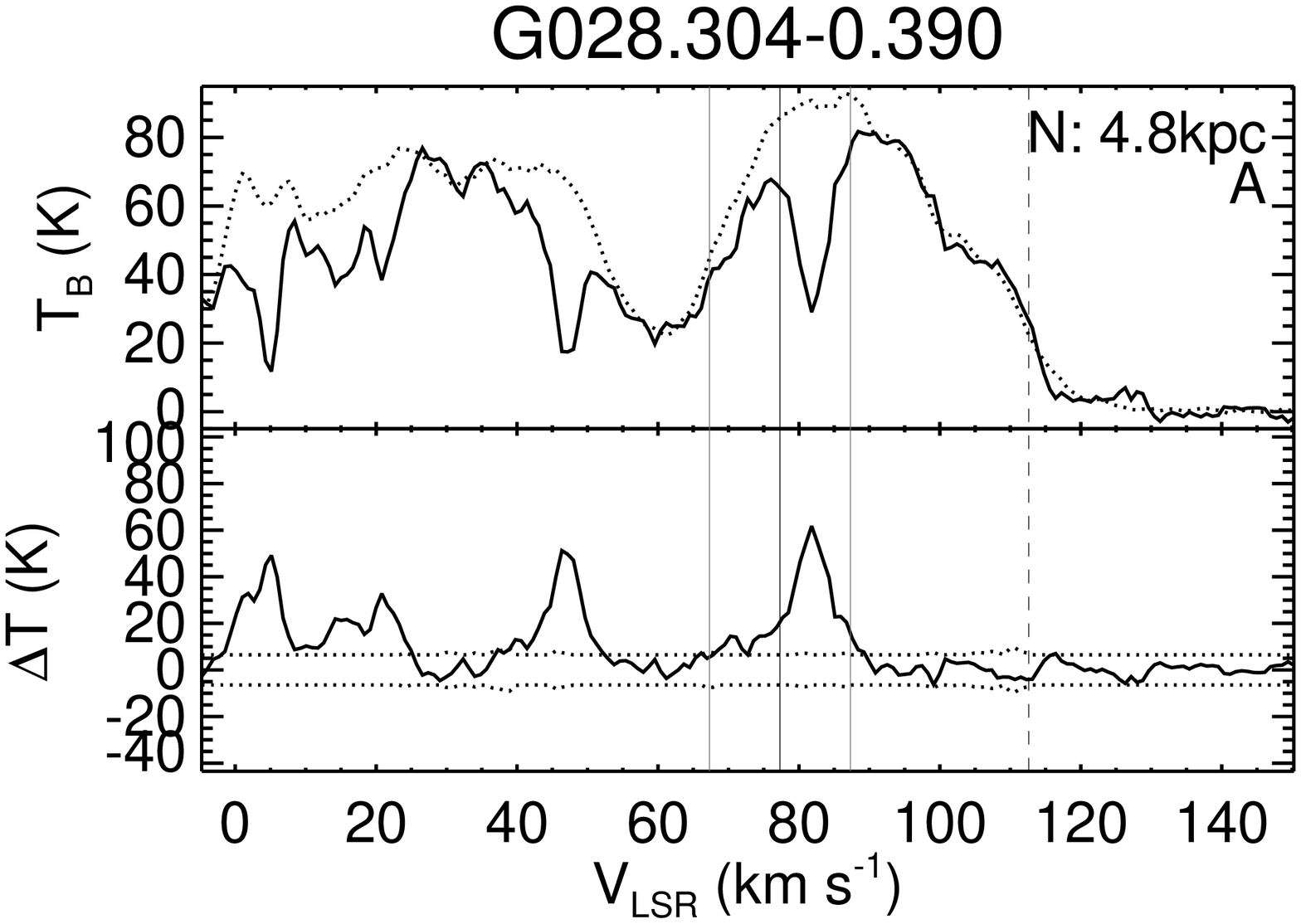} 
\caption{Example difference spectra for the same HRDS sources shown in
  Figure~\ref{fig:apertures}.  The top panel of each plot shows the
  on-- (solid line) and off-- (dotted line) source average
  \hi\ spectra.  The bottom panel of each plot shows the difference
  between the off-- and on--source spectra (solid line) and our error
  estimates (dotted lines; see \S\ref{sec:method}).  The three solid vertical lines mark the
  RRL velocity and $\pm$10 \kms\ of the RRL velocity, and the dashed
  vertical line marks the tangent point velocity.  We assigned each
  source a qualitative quality factor (QF) of ``A,'' ``B,'' or ``C''
  (see text).  Clockwise from top-left are a far source of QF A, a far
  source of QF B, a near source of QF A, and a tangent point source
  (where no KDA resolution would be possible based on the difference
  spectrum, and no QF is assigned).  In the top right corner of each
  plot we give the KDA resolution (``N,'', ``F,'', or ``T'' for near,
  far, or tangent point), followed by the assigned distance in kpc.
  Below this we print the QF.}

\label{fig:spectra}
\end{figure}

We visually examine the difference spectra to determine the maximum
velocity of \hi\ absorption for each source, and thus the resolution
of the KDA.  We show example spectra in Figure~\ref{fig:spectra} for
the same four sources displayed in Figure~\ref{fig:apertures}.  The
top plot in each of the panels of Figure~\ref{fig:spectra} is the
on--source (solid line) and off--source (dotted line) average
\hi\ spectra.  The bottom plot is the difference spectrum.  The RRL
velocity from Paper~II is marked with a solid vertical line, as are
the velocities $\pm10$\,\kms\ of the RRL velocity.  The vertical
dashed line shows the tangent point velocity as calculated with the
\citet{brand86} rotation curve.  The dotted lines in the bottom panel
show the error estimates, the maximum at each spectral channel of $5\sigma_{\rm rms}$ and
  $\sigma_T(v)$.

\begin{figure}
\centering
\includegraphics[width=3 in]{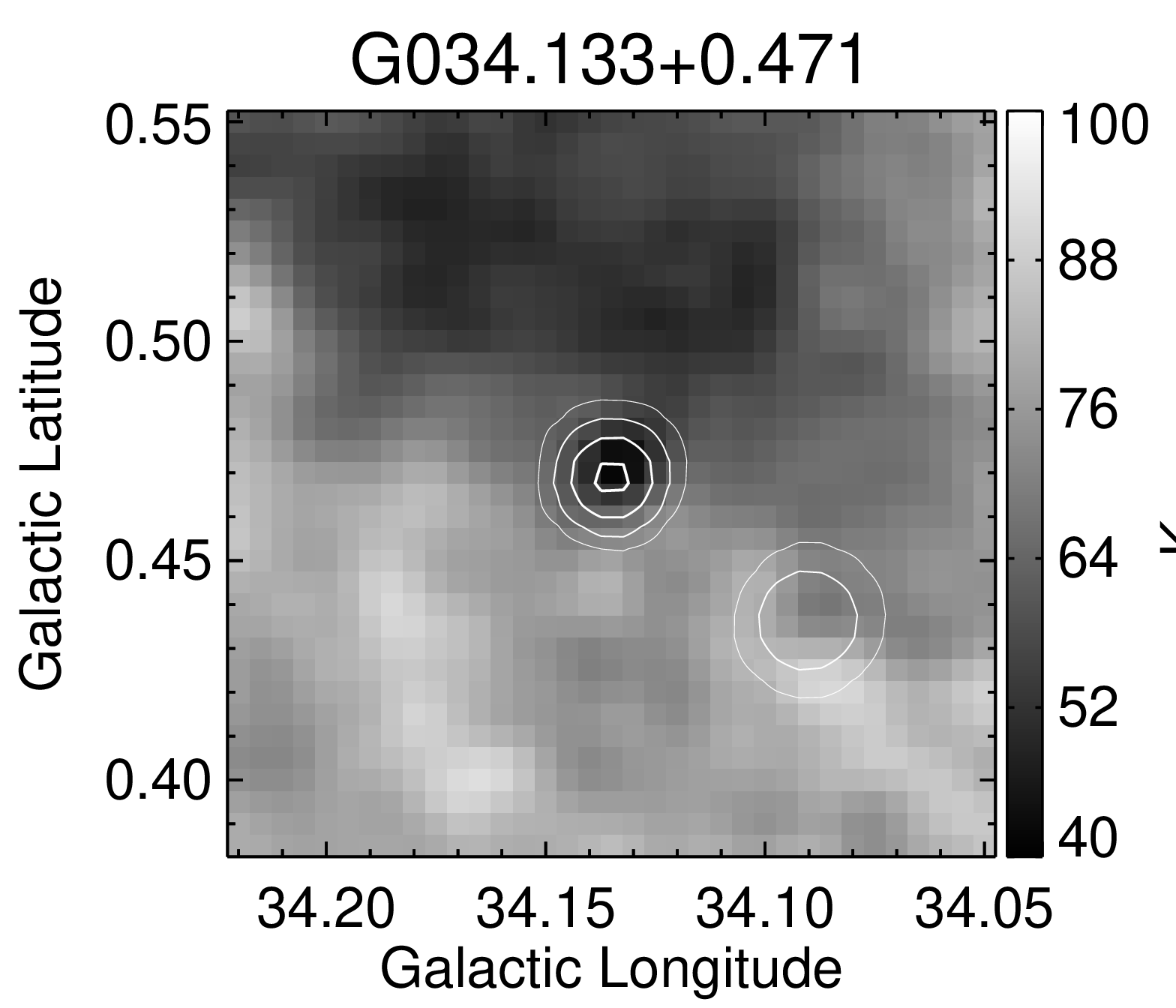}
\includegraphics[width=3 in]{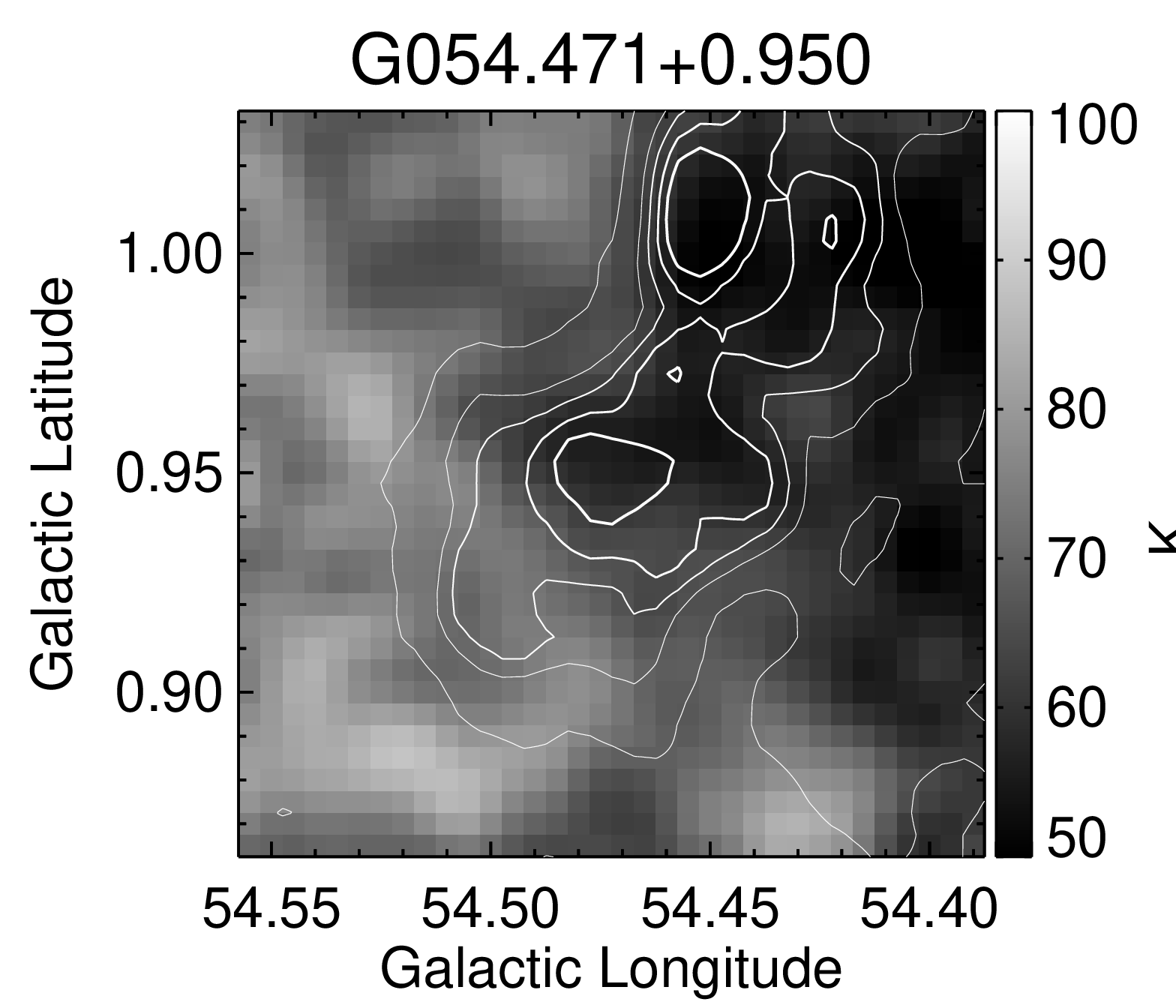}
\includegraphics[width=3 in]{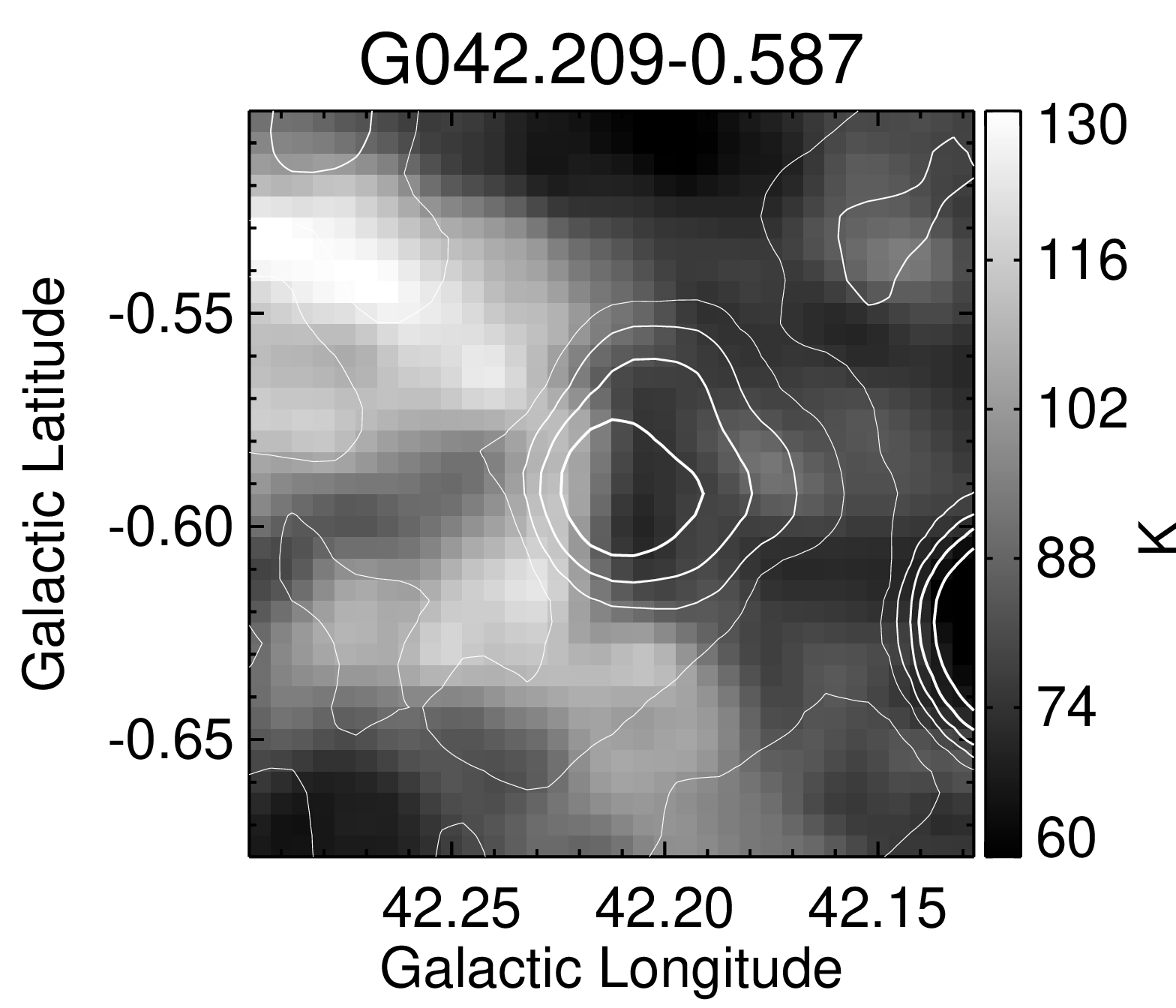}
\includegraphics[width=3 in]{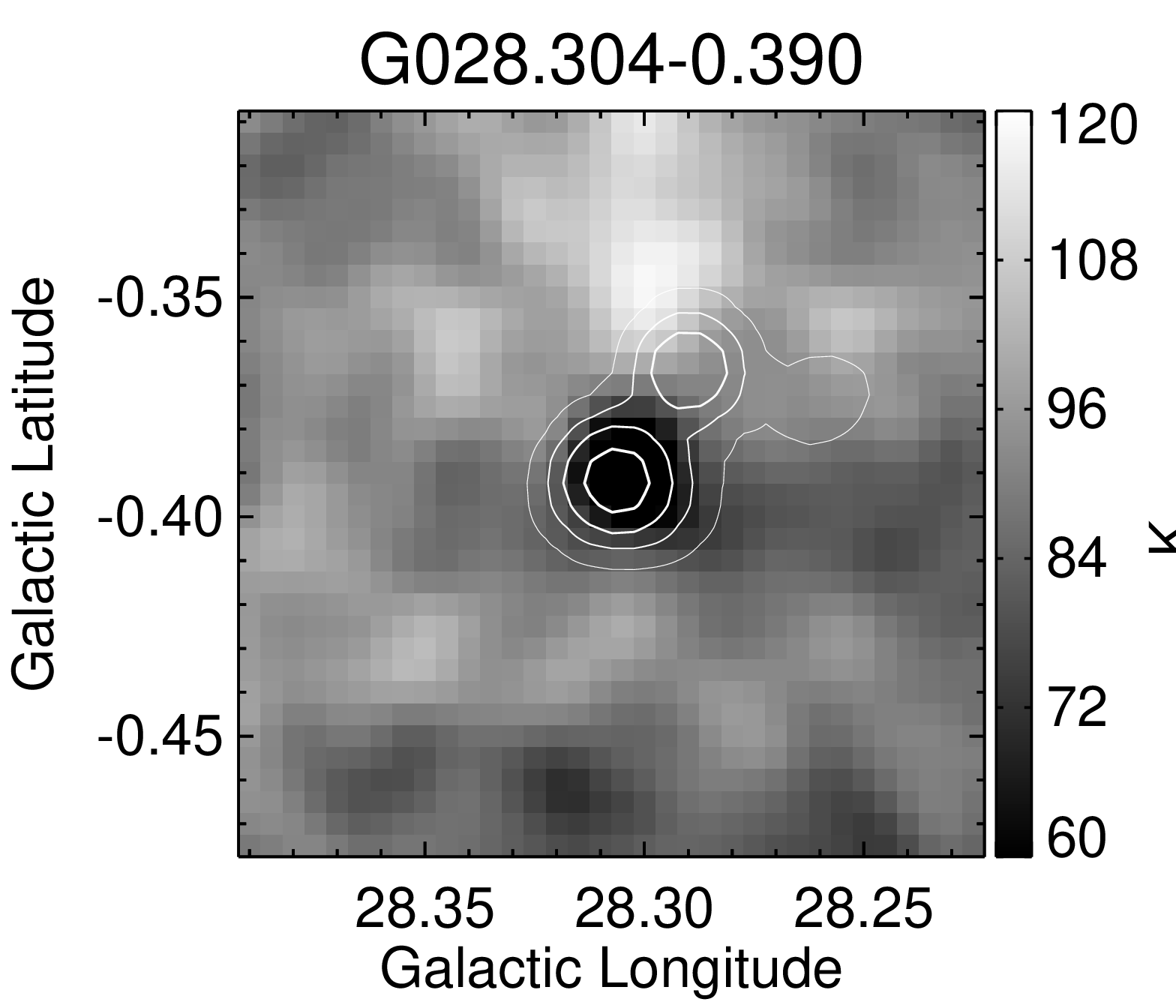}

\caption{Example \hi\ channel maps at the velocity of maximum absorption
  for the same sources shown in Figure~\ref{fig:apertures}.  The contours
  are 21\,cm VGPS continuum emission.  The \hii\ regions are at the
  center of their respective panels.  The morphological agreement
  between the \hi\ absorption and the radio continuum emission
  demonstrates that the absorption signals are likely real (although the situation is slightly ambiguous for G042.209$-$0.587, which is located at the tangent point).}

\label{fig:channelmaps}
\end{figure}

As AB did, we verify all identified features of maximum absorption
using VGPS \lb\ \hi\ channel maps at the velocity of maximum detected
absorption.  If there is no absorption seen in the \lb\ image with a
similar morphology to the continuum emission of the HRDS source, we
regard this absorption feature as spurious and repeat the analysis
for an absorption feature detected at a lower velocity.  If there are
no lower velocities with detected absorption, we cannot resolve the
KDA.  This step is very important because \hi\ self-absorption, the
absorption of the emission from warm background \hi\ by cold foreground \hi\ at
the same velocity \citep[see][]{knapp74, liszt81, jackson02,
  gibson05}, can mimic \hiea.  In other words, not all absorption
signals detected in the difference spectrum are caused by the
continuum emission of the \hii\ region.  If the morphology of the
absorption signal does not match that of the \hii\ region continuum
emission, this is a sign that the absorption signal in question is not
caused by \hiea.  Example channel map plots are shown in
Figure~\ref{fig:channelmaps} for the same four regions displayed in
Figures~\ref{fig:apertures} and \ref{fig:spectra}.


We assign for each source a quality factor (QF) based on our
confidence that the KDA was resolved correctly.  This qualitative
factor takes into account the number of absorption signals detected,
the strength of said signals, the distance from the source to the
tangent point, and the morphological agreement between the absorption
and the radio continuum emission from the source.  As AB did, the QF
can have a value of ``A'' or ``B'' for sources with resolved KDAs, or
``C'' for sources too faint for a KDA resolution.  Sources for which
we assign the tangent point distance have no QF.  QF~A sources are our
most confident determinations and are characterized by strong
absorption well above the noise estimates and a good
  morphological \lb\ match between the absorption signal and the
  source radio continuum emission.  QF~A sources at the far distance
generally have multiple absorption features between the source
velocity and the tangent point velocity.  QF~B sources have weaker
absorption and the KDA resolution is frequently based on a single
absorption feature.  The morphological agreement between the
  absorption and the source radio continuum emission may be poor for a
  QF~B source.  Sources whose velocity is close to that of the
tangent point velocity more frequently have B QF designations.  
  We encourage other researchers who wish to consider only the most
  robust KDA resolutions to use only the QF~A distances.


\section{Results}
We derive kinematic distances to 149 of 182 HRDS sources.  Excluding
sources for which we assigned the tangent point distance and
negative-velocity sources for which there is no KDA, we were able to
resolve the KDA for 85 of 118 HRDS \hii\ regions (72\%).  Although
they are fainter on average than the \hii\ regions in AB, the small
size of the HRDS nebulae allows us to resolve the KDA for a high
percentage of sources.  For small sources, we may define on-- and
off--source apertures near to each other in angle, and the two apertures
therefore better sample the same gas along the line of sight.  The sources for which we were unable to resolve
the KDA have no absorption above our error estimates whose spatial morphology matches that
of the source radio continuum emission, and thus no
distance assignment can be made with confidence.

We give the KDA results in Table~\ref{tab:data}, which lists for
each source its name, Galactic longitude and latitude, LSR velocity
from Paper~II, maximum velocity of detected \hi\ absorption, tangent
point velocity, near and far distances, KDA resolution, QF, derived
heliocentric distance, calculated uncertainties in the derived distance,
Galactocentric radius, and distance from the Galactic plane, $z$.  We calculate all
kinematic distances and tangent point velocities using the
\citet[hereafter B86]{brand86} rotation curve.  We compute the
distance uncertainties from our estimates of the uncertainties caused
by the choice of rotation curve model, non-circular velocities, and a
change to the circular rotation speed of the LSR (see \S\ref{sec:uncertainties}).

\begin{figure} \centering
\includegraphics[width=5.5 in]{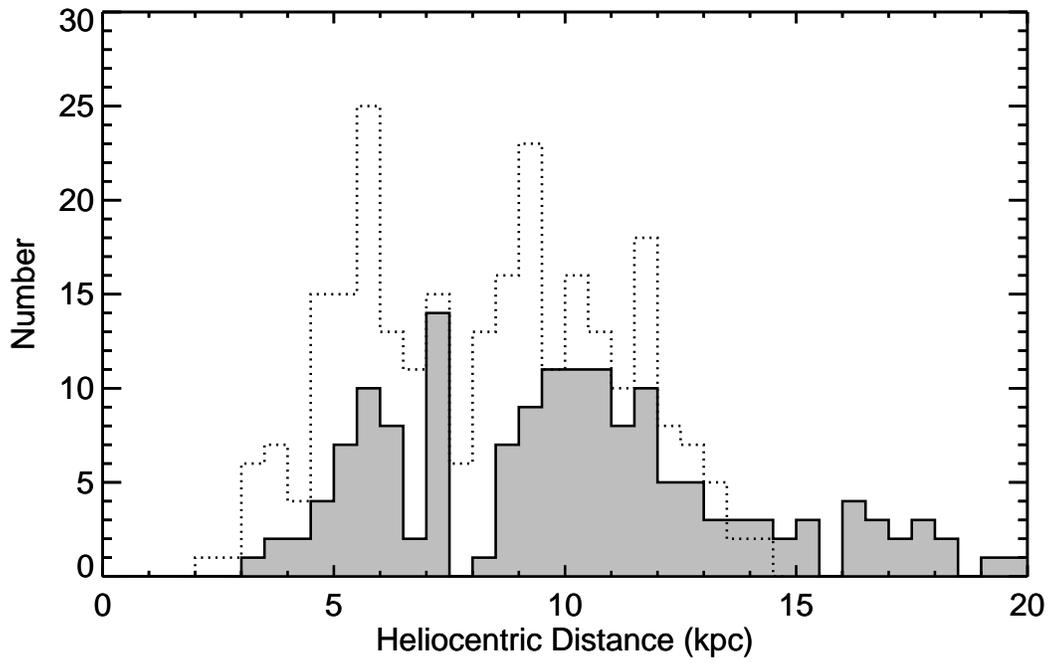}

\caption{Distribution of Heliocentric distances HRDS (gray filled) and
  the AB sample (dotted line).  The HRDS nebulae are on average more
  distant, and in fact the two samples are statistically distinct.}

\label{fig:distances}
\end{figure}

HRDS sources are on average more distant than \hii\ regions known
previously.  The average distance for the HRDS nebulae is 10.1\,kpc
whereas the average distance in the AB sample is 8.4\,kpc.  AB
used the rotation curve of \citet{mcclure07}.  We have
recomputed kinematic distances and Galacticentric radii for the
\hii\ regions in AB using the B86 rotation curve.  We use these
recomputed distances for all analyses involving the \hii\ regions from
AB.  Thus all analyses discussed here are based on kinematic distances derived using the
same B86 rotation curve.  A
Kolmagorov-Smirnov (K-S) test shows that the Heliocentric distances to
the objects in the two samples are statistically distinct.  In
Figure~\ref{fig:distances} we show the distribution of heliocentric
distances for the HRDS (gray filled) and AB samples (dotted line).
Figure~\ref{fig:distances} shows that the HRDS nebulae are on average
more distant from the Sun than the AB sample, and that almost nothing
was known about the \hii\ region population beyond 15\,kpc from the
Sun in this zone of the Galaxy.  
The relative lack of HRDS sources
within 7\,kpc of the Sun indicates that the sample of \hii\ regions
close to the Sun was more complete prior to the HRDS.  The two samples
share a similar distribution from $10-15$\,kpc.

In addition to being on average more distant, the HRDS sample contains the most distant known \hii\ regions.  There are
19 HRDS regions whose kinematic distances derived here are greater
than 15\,kpc from the Sun, and nine with distances greater than
17\,kpc.  Prior to the HRDS, there were six known regions with
distances calculated with the B86 curve greater than 15\,kpc and just
three with distances greater than 17\,kpc.  In this tally we used the
``known'' sample from Paper~II, restricted the range to $70\degree >
\ell > -70\degree$, and excluded sources within 15$\degree$ of the
Galactic center.  The most distant regions detected in the HRDS are
G031.727+0.698 and G032.928+0.607, which have heliocentric distances
of 19.7\,kpc and 19.2\,kpc, respectively.  Of the \hii\ regions known
prior to the HRDS, S83 \citep{sharpless53} located at \lb = (55.114,
+2.422), has the largest distance from the Sun.  Its RRL velocity of
$-$81.5 \kms\ \citep{lockman89} places it 19.4\,kpc from the Sun
according to the B86 curve.
This region is well off the Galactic plane.  Vertical derivatives in
rotational velocities are not taken into account in the B86 curve
\citep[although they are in other curves, e.g.,][]{levine08} and
therefore for sources well off the Galactic plane the conversion from
radial velocity to distance is more uncertain.  While S83 is sure to
be extremely distant, its distance derived with the B86 curve has
larger error bars than a comparable source in the Galactic plane.


Nearly all HRDS sources are at the far kinematic distance: 61\% of
HRDS sources are located at their far distance, 31\% are at the tangent
point distance, and only 7\% are at their near distance (excluding
negative-velocity sources for which there is no distance ambiguity).
Excluding sources for which we assign the tangent point distance, 89\%
are at the far kinematic distance and only 11\% are at the near
kinematic distance.  This implies that the small angular size of the
HRDS nebulae (see Paper~II) is due to their large distance from the
Sun and not to a small physical size.  For comparison, AB assigned the
far distance to approximately two thirds of their sample, and the near
distance to one third (excluding tangent point distance sources).

If \hii\ regions were evenly distributed out to a Galactocentric distance
of 8.5\,kpc, for the longitude limits of the present study we would
expect to find two-thirds of all \hii\ regions at the far distance and
one-third at the near distance, as AB found.  The combined AB and HRDS
sample has 73\% of all sources at the far distance and 27\% at the
near distance (again excluding negative-velocity sources and source at
the tangent point distance).
That we have
such a large population at the far distance suggests the
sample is complete to the same degree for near- and far-distance
\hii\ regions out to the Solar orbit.

``Bubble'' \hii\ regions that have an annulus of emission at 8.0\,\micron\
surrounding the ionized gas are not at the near distance as was assumed by
\citet{churchwell06}.  Paper~II classified all HRDS targets based on
their 8.0\,\micron\ morphology.  Since there are so few near-distance
sources, it is not surprising that there is little difference in mean
heliocentric distance between the classifications -- all average
$\sim\!10$\,kpc.  We derive distances to 55 Galactic bubbles (Paper~II
classifications of ``Bubble'', ``Bipolar Bubble'', ``Partial Bubble'',
and ``Irregular Bubble'').  Of these, 42 are at the far distance, 10
are at the tangent point distance, and only three are at the near
distance.  The average heliocentric distance for these 55 sources is
10.7\,kpc; it is 11.1\,kpc for the ``Bubble'' classification alone.

\begin{figure} \centering
\includegraphics[width=5.5 in]{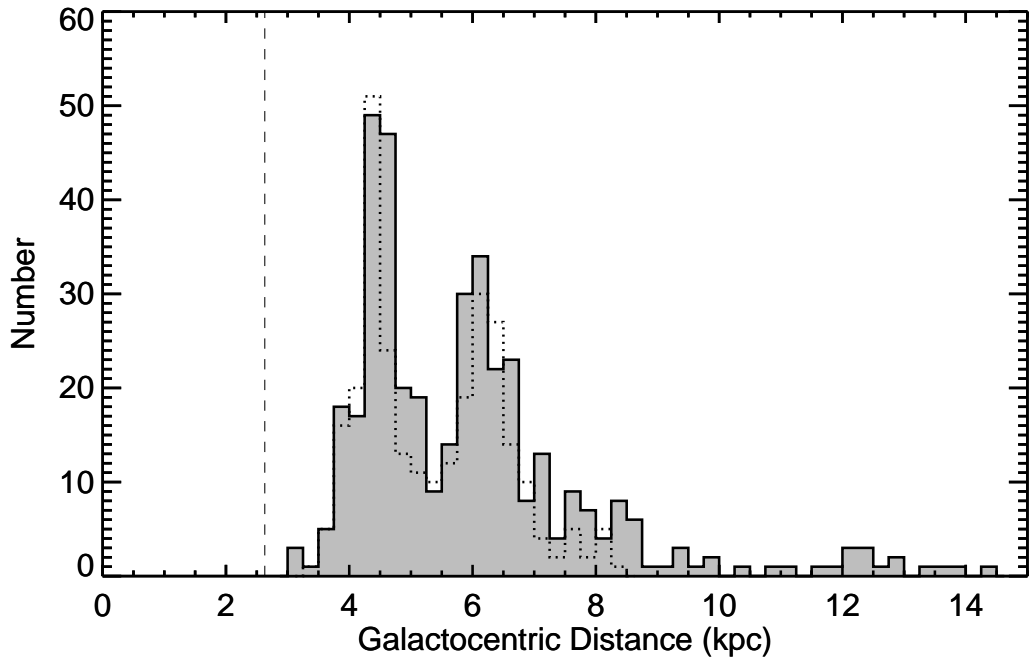}

\caption{Distribution of Galactocentric radii for the HRDS (gray
  filled) and the AB sample (dotted line).  The two samples are
  statistically similar.  The vertical dashed line shows the minimum
  Galactocentric radius sampled by the present study.  The peaks at
  4.25 and 6.0\,kpc are only about 1\,kpc wide, for both samples.  }

\label{fig:rgal}
\end{figure}

In Figure~\ref{fig:rgal} we show the Galactocentric radius
distrubution for the HRDS (gray filled) and AB nebulae (dotted
line)\footnote{This figure is similar to that of Paper~I (their
  Figure~3) but is restricted here to the range of the current
  study.}.  There are two obvious peaks at 4.25\,kpc and 6.0\,kpc in
both distributions.  A K-S test shows that the two samples are not
statistically distinct.  Many previous authors have found peaks in
tracers of star formation at these Galactocentric radii over similar
areas of the Galactic plane: \citet{mezger70}, \citet{lockman79}, \citet{downes80}, and AB for \hii\ regions,
\citet{schlingman11} for spectroscopic observations of sub-mm clumps identified in the Bolocam
Galactic Plane Survey, and (less clearly) by \citet{roman-duval10} for
\cor\ clouds identified by \citet{rathborne09} in the Galactic Ring
Survey \citep{jackson06}.  In Figure~\ref{fig:rgal} these peaks are extremely narrow, just
1\,kpc FWHM when modeled with a Gaussian (see Paper~I), and are present with the same properties for both the AB
and the HRDS samples, despite the different distances probed by the two
studies.  That the HRDS Galactocentric radius distribution is
statistically similar to that of the previously known \hii\ regions
suggests that the HRDS nebulae are not a new population of
\hii\ region but rather are just fainter versions of \hii\ regions
previously identified.

\begin{figure}
\centering
\includegraphics[width=6.5 in]{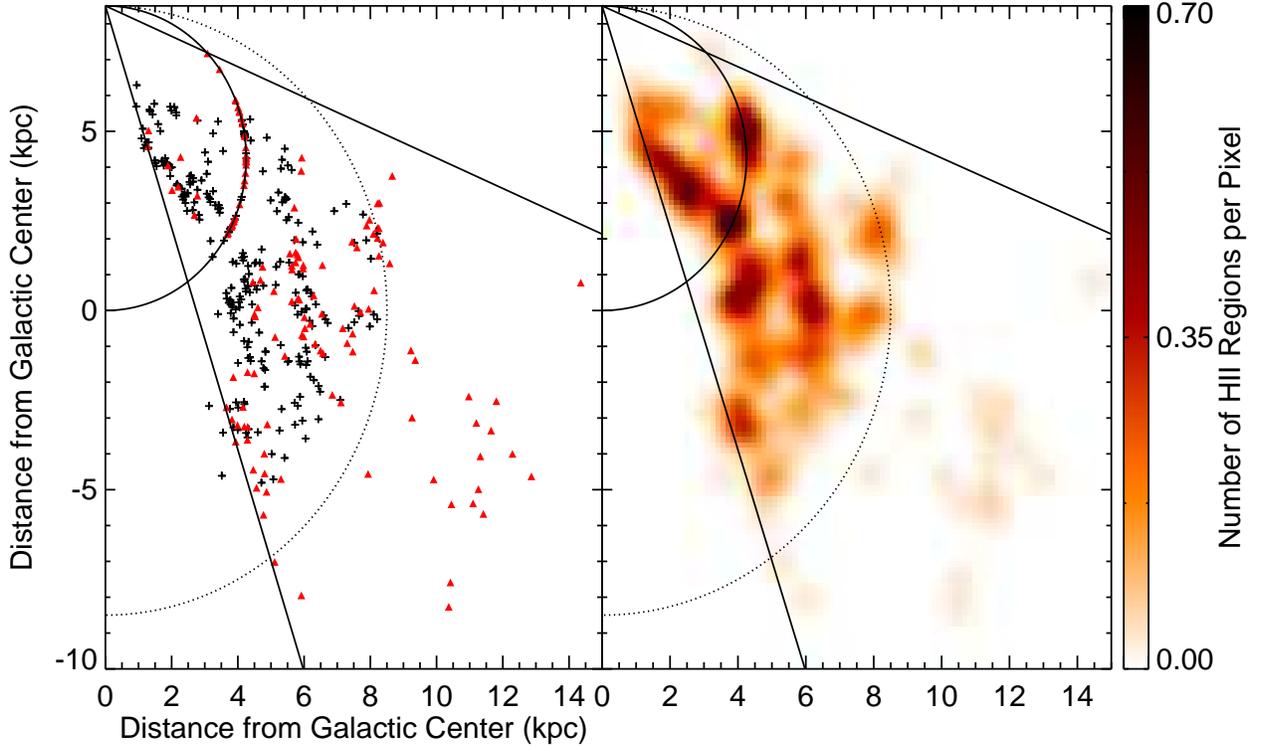}

\caption{Galactic distribution of \hii\ regions.  The Sun is located
  in the upper left corner and the Galactic center is at (0,\,0).  The
  plot contains both HRDS sources (triangles) and also \hii\ regions
  known prior to the HRDS from AB (crosses).  The straight solid lines
  show the longitude range of the present study, $17\fdg9$ to
  $67\arcdeg$ (the longitude range of AB is different).  The solid
  half-circle shows the tangent-point distance and the dotted half
  circle shows the Solar orbit.  The left panel shows the positions
  for all regions with assigned distances.  In the right panel, we
  binned these positions into $0.15 \times 0.15$\,\kpc\ pixels, and
  smoothed the resultant image with a $5 \times 5$ pixel Gaussian
  filter.  The semicircular arc-segments correspond to the peaks in
  the Galactocentric radius distribution seen in
  Figure~\ref{fig:rgal}.}

\label{fig:faceon}
\end{figure}

We show in Figure~\ref{fig:faceon} the face-on distribution of the 153
HRDS regions for which we derive kinematic distances, as well as the
261 previously known \hii\ regions with derived distances from AB.  In the left panel of Figure~\ref{fig:faceon}, we plot HRDS
sources as triangles and the sources from AB as crosses.  The Sun is
located in the upper left corner and the Galactic center is located at
(0, 0).  In the right panel, we binned the data into 0.15\,kpc pixels
and smoothed the resultant distribution with a $5 \times 5$\,pixel
Gaussian filter. The solid half-circle shows the tangent point
locations and the dotted half-circle shows the Solar orbit.  The solid
lines show the longitude range of the present study.

Figure \ref{fig:faceon} shows signs of Galactic structure traced by
\hii\ regions.  There are two circular arc segments centered at the
Galactic Center with mean Galactocentric radii of 4.25 and 6.0\,kpc;
these map directly to the two peaks identified in
Figure~\ref{fig:rgal}.  These locations are near where the Scutum and
Sagittarius arm are thought to be; for example, large streaming
motions are found at these Galactocentric radii \citep{mcclure07}.  As
have many previous authors \citep[][AB]{burtongordon76, lockman81} we
find a dearth of \hii\ regions within 3.5\,kpc of the Galactic center,
although this region of the Galaxy is not well-sampled by the present study.
AB hypothesized that this feature is due to a Galactic bar of
half-length 4~kpc \citep[see][]{benjamin05}.  The extreme distances of
the negative-velocity sources are clearly visible.  It is unclear,
however, whether their loose grouping is physical or due to
difficulties applying a rotation curve model.  Aside from the greater
distances, there is little difference between the distribution of HRDS
sources and that of AB.

\section{Uncertainties in Kinematic Distances\label{sec:uncertainties}}
There are many possible sources of uncertainty when computing
kinematic distances.  Errors in kinematic distances affect the
interpretation of Galactic structure traced with \hii\ regions,
including derived electron temperature gradients
\citep[e.g.,][]{balser11}.  Here we consider three sources of
kinematic distance uncertainty.  First, there is uncertainty based on
the choice of rotation curve model.  Secondly, large-scale non-circular
motions caused by streaming motions along spiral arms are generally not accounted for in
axisymmetric circular rotation curve models, and this omission may
cause significant uncertainty in derived distances.  Finally, the
standard parameters used when computing distances from a rotation
curve (the Sun's distance from the Galactic center and the Solar
orbital speed) may need modification from the IAU standard values
\citep[e.g.,][]{reid09}.  Throughout, we compare all sources of
uncertainty to the distances derived using the rotation curve of
B86.

The full details of our analysis can be found in
Appendix~\ref{sec:appenb}.  Briefly, we compute for a grid of
\lv\ locii the difference in distance between that of the B86 curve,
and the distance found after accounting for a given source of uncertainty.  We compute
these distance differences separately for each of the three sources of
uncertainty we consider.

\begin{figure}
\centering
\includegraphics[width=2.9 in]{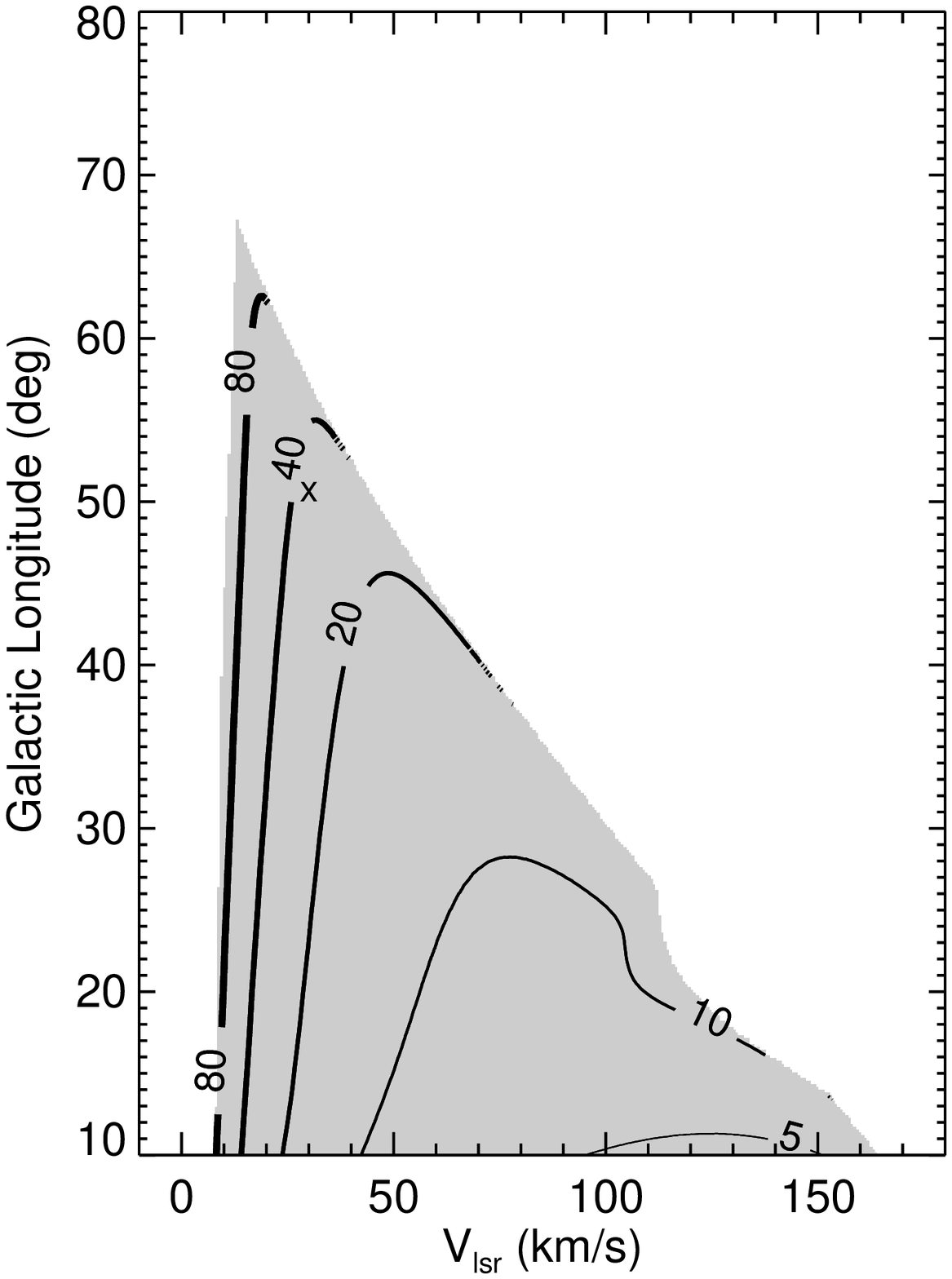}
\includegraphics[width=2.9 in]{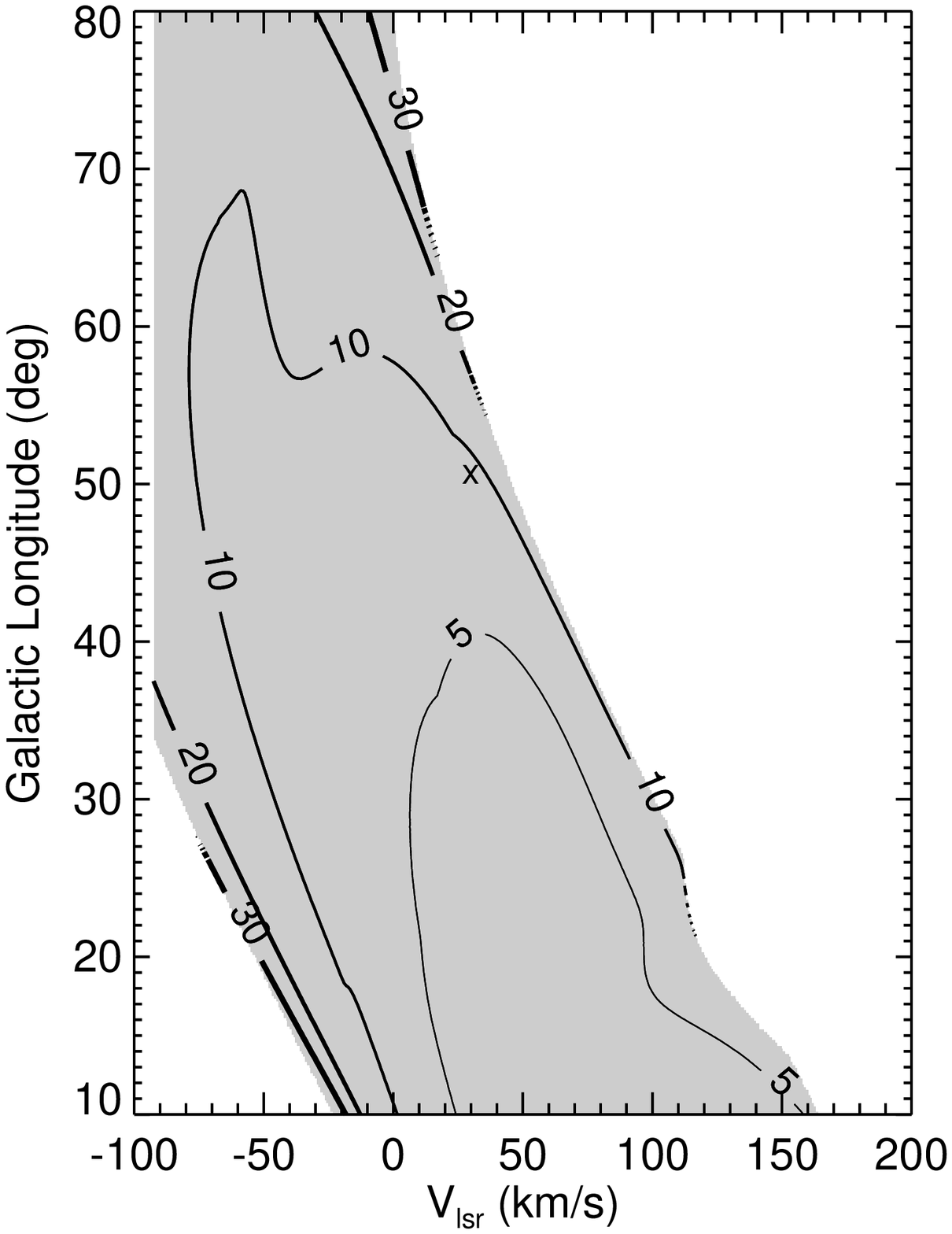}

\caption{The total uncertainties associated with the choice of
  rotation curve, the effect of non-circular motions, and changing the
  Solar rotation speed for near distances (left panel) and far
  distances (right panel).  We mark the example position mentioned in
  the text, \lv$=(50\degree, 30\,\kms)$, with an ``x''.}

\label{fig:big_nearfar}
\end{figure}

We add the effect of these three sources of distance uncertainty
in quadrature for each \lv\ locus to compute a total
uncertainty.\footnote{Differences in rotation curve models arise
  in part from the other sources of uncertainty considered here and therefore
the three sources of uncertainty are not independent.  For example, the \citet{clemens85}
  curve fits for streaming motions, which causes some of the
  ``waviness'' seen in Figure~\ref{fig:rgal_theta}.}  
We divide this total uncertainty by the
distances derived using the B86 curve for each \lv\ locus to
compute a ``percentage uncertainty.''  We show this
percentage uncertainty in the near (left panel) and far (right panel)
distances in Figure~\ref{fig:big_nearfar}.
Each \lv\ locus in
Figure~\ref{fig:big_nearfar} has a corresponding uncertainty in both
panels.  For example, \lv$=(50\degree, 30\,\kms)$ has an uncertainty
of 38\% for the near distance and 9\% for the far distance; this
locus is marked in Figure~\ref{fig:big_nearfar} with an ``x''.  

\begin{figure}
\centering
\includegraphics[width=3.5 in]{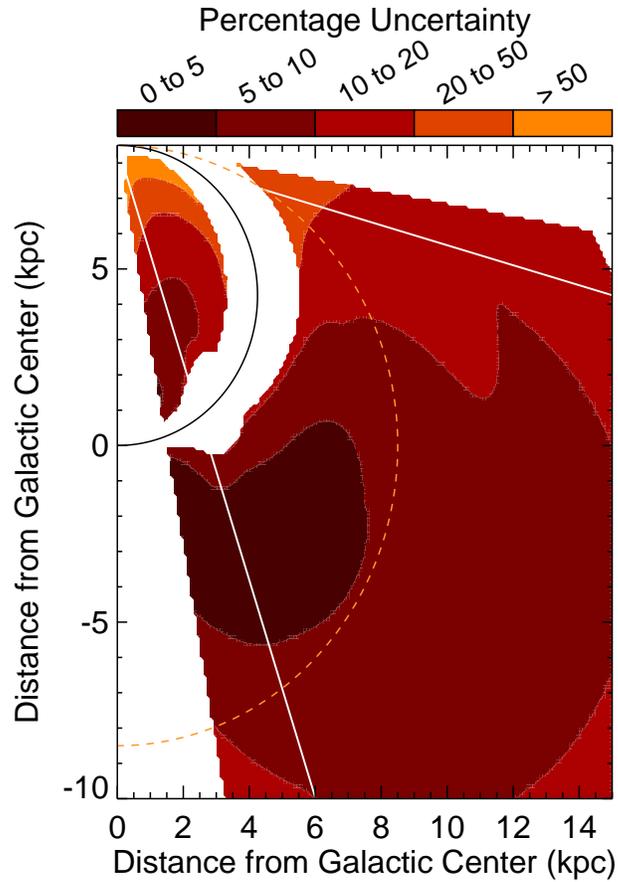}

\caption{Face-on map of the total percentage uncertainty in kinematic
  distances caused by the choice of rotation curve, non-circular motions
  of 7\,\kms, and by changing the Solar circular rotation speed to
  250\,\kms.  The meaning of the curves and lines are
  as in Figure~\ref{fig:faceon_three}.}

\label{fig:big_faceon}
\end{figure}

We transform the data of Figure~\ref{fig:big_nearfar} into the face-on
plot of distance uncertainties shown in Figure~\ref{fig:big_faceon}.
To construct this figure, we find for each \lv\ locus the corresponding distance using
the B86 curve.  We then use the corresponding percentage uncertainty
from Figure~\ref{fig:big_nearfar} at each \lv\ locus for the value
in the face-on map.  The white holes in Figure~\ref{fig:big_faceon}
correspond to \lv\ locii that are not defined for all trials of the
error analysis (see Appendix~\ref{sec:appenb}).  Only $\sim 20\%$ of the \lv\ locii
in Figure~\ref{fig:big_faceon} have uncertainties $\leq 5\%$, but over
60\% of the locii have uncertainties $\leq10\%$ and $\sim\,90\%$
of the locii have uncertainties $\leq20\%$.  Uncertainties are
greater near the Sun and at higher Galactic longitudes.

What effect do these uncertainties have on the Galactic distribution of \hii\ regions?
For each source in the combined HRDS and previously
known (from AB) samples we compute the difference in the B86 distance
caused by three effects: 1)~when the \citet{clemens85} curve is used 2)~with
non-circular motions of maximum 7\,\kms\ and minimum $-7$\,\kms, drawn
randomly from a uniform distribution; and 3)~when the Solar rotation speed
is changed to 250\,\kms.  (Here we have scaled the
\citet{clemens85} curve so that it has a Solar rotation speed of 220\,\kms, instead of the 250\,\kms\ value.)  For each of these three
sources of uncertainty, we compute the difference in derived distance
from that calculated with the B86 curve, preserving the sign of the
difference.  We add these three differences to the B86 distance to
create an adjusted distance.  An alternate method would be to add
differences in quadrature, as we did when estimating the
uncertainties.  Since the differences do not always have the same sign
(they do not, for example, always increase the distance computed with
the B86 curve), our method estimates what effect these distance
uncertainties may have on the Galactic distribution of \hii\ regions
and is applicable for all Galactic locations.  We stress that
  this is the worse-case scenario where we have assumed that both the
  rotation curve and also the Solar circular rotation speed are
  incorrect, and the rotation curve model does not account for a
  change in Solar rotation speed.  

We find that the sources of uncertainty investigated here have a
relatively minor effect on \hii\ region distances.  The median
absolute differences in distance for our combined sample of
\hii\ regions are 0.2\,kpc, 0.2\,kpc, and 0.4\,kpc for changes to the
rotation curve model, non-circular motions, and the Solar rotation
speed, respectively.  The median percentage differences are
respectively 2\%, 4\%, and 4\%.  The combined median absolute
difference is 0.5\,kpc, or 5\%.  \citet{gomez06} found a similar
result using a simulation of the velocity field of the Galaxy.  He
found that the difference between the distance inferred from a
rotation curve and the true distance is $<0.5$\,kpc for the majority
of the Galactic disk.  The median distance between the near and the
far distances calculated using the B86 curve for our combined sample
of \hii\ regions is 6.0\,kpc, after excluding sources at the tangent
point and those beyond the Solar orbit.  {\it Thus, errors in
  kinematic distances are very small relative to the uncertainties associated
  with the KDA.}

\begin{figure}
\centering
\includegraphics[width=6.0 in]{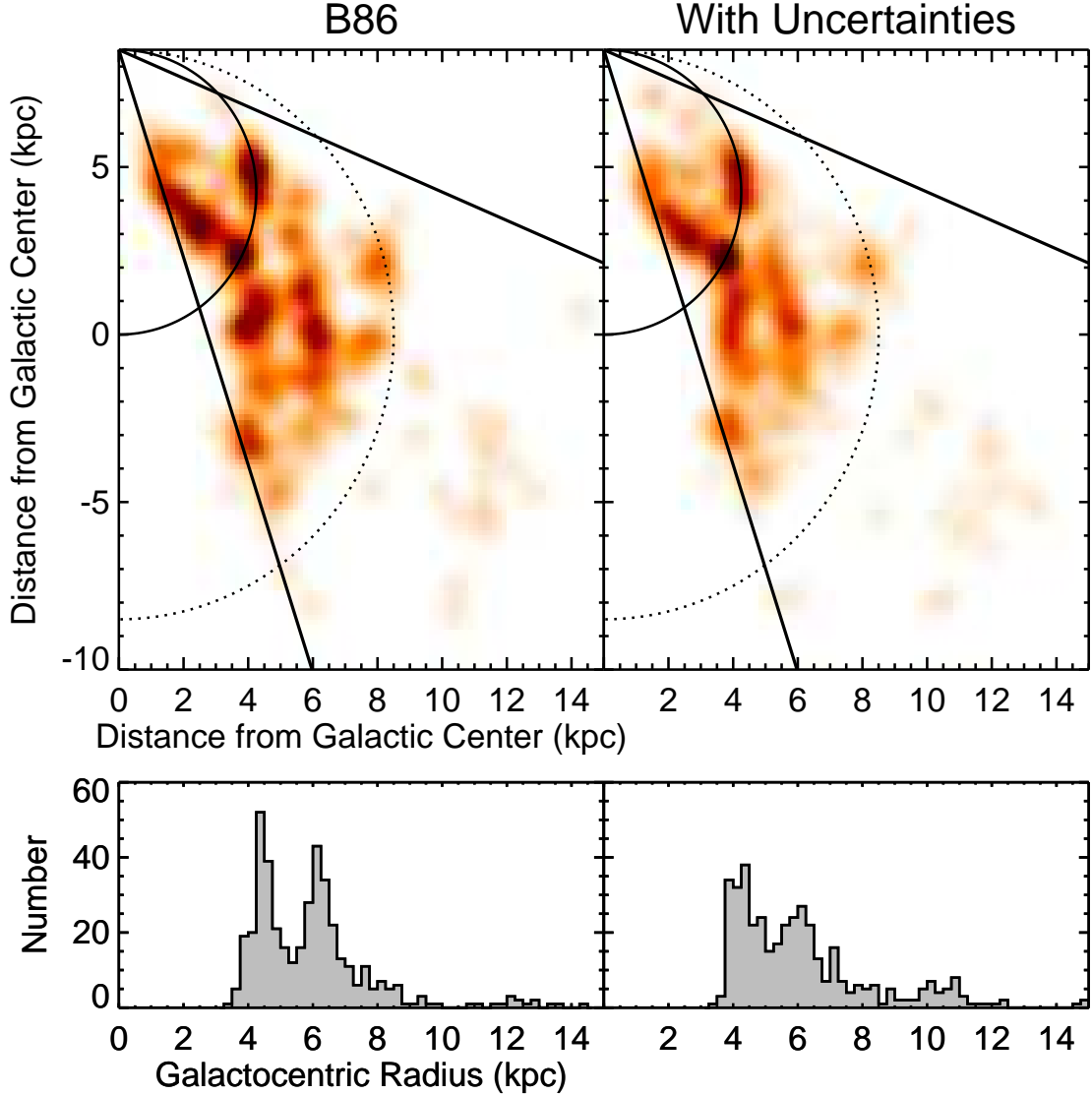}

\caption{The effect of distance uncertainties on
  Figures~\ref{fig:rgal} and \ref{fig:faceon}.  In the top row we show
  the smoothed face-on plot of the HRDS and previously known
  \hii\ regions; in the top left panel we calculate the distances
  using the B86 curve and in the top right panel we have accounted for
  uncertainties caused by the choice of rotation curve, non-circular
  motions, and a change to the Solar circular rotation speed.  The
  lines in the top panels are as in Figures~\ref{fig:faceon}.  The
  bottom row shows the distribution of Galactocentric radii for the
  B86 curve (bottom left panel), and after accounting for these
  sources of distance uncertainty (bottom right panel), for the same
  sources shown in the top row.  The basic structures are largely
  unchanged.}

\label{fig:faceon_compare}
\end{figure}

\begin{figure}
\centering
\includegraphics[width=5 in]{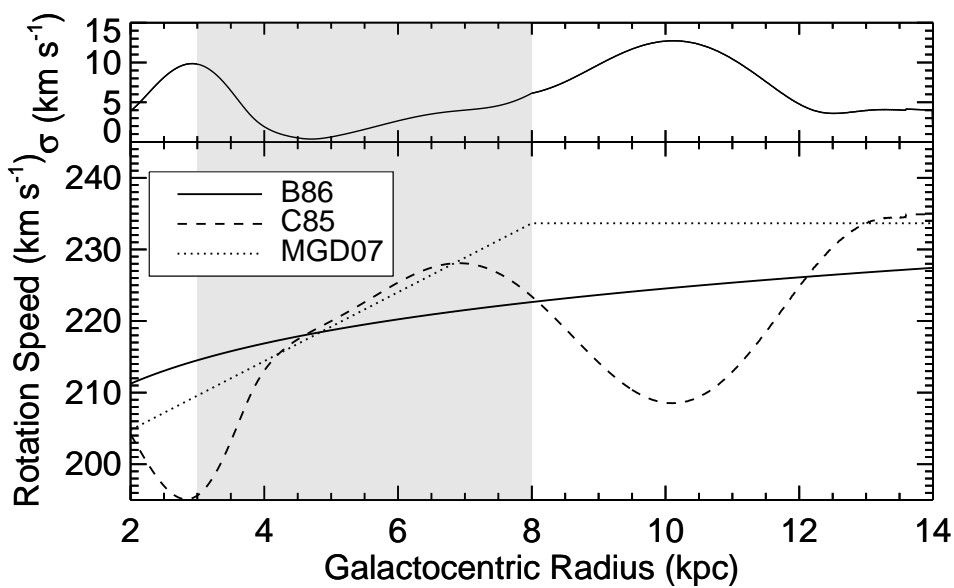}

\caption{The circular rotation speed versus Galactocentric radius for
  the B86 curve (solid line), the C85 curve (dashed line), and the
  MGD07 curve (dotted line).  In the upper panel, we show the standard
  deviation between the three curves.  The shaded region shows the
  range of Galactocentric radii over which the MGD07 curve is defined,
  3--8\,kpc.  We extrapolate the MGD07 curve below 3\,kpc and assume a
  flat rotation curve above 8\,kpc.}

\label{fig:rgal_theta}
\end{figure}

In Figure~\ref{fig:faceon_compare} we show graphically the effect of
the above sources of uncertainty on our derived Galactic structure
results, using the combined HRDS and previously known \hii\ region
samples.  The top two panels in Figure~\ref{fig:faceon_compare} have
the same format as Figure~\ref{fig:faceon}.  The top left panel of
Figure~\ref{fig:faceon_compare} is in fact identical to the right
panel of Figure~\ref{fig:faceon}, where distances are calculated using
the B86 curve, and the top right panel of
Figure~\ref{fig:faceon_compare} shows the adjusted distances
after applying the uncertainties discussed previously.  The bottom two
panels show the Galactocentric radius distribution; the bottom left
panel has Galactocentric radii from the B86 curve for \hii\ regions
with derived distances and the bottom right panel has adjusted
Galactocentric radii after examining the distance uncertainties.

{\it While the distance calculated for individual H{\footnotesize II}
  regions may be uncertain by 10\%, the overall distribution in this
  zone of the Galaxy is little effected by the uncertainties
  investigated here.}  The basic findings of this work are unchanged
after accounting for these sources of uncertainty.  We still find a
dearth of \hii\ regions within 3.5\,kpc of the Galactic center and
there are still concentrations of \hii\ regions near 4.25\,kpc and
6.0\,kpc.  The width of these peaks in Galactocentric radius has
grown, and their height has decreased, after factoring in the sources
of uncertainty.  The overall face-on picture is visually similar.

\section{Conclusions}
Using the \hi\ Emission/Absorption method, we resolved the kinematic
distance ambiguity and derived kinematic distances for 149 of 182
(82\%) \hii\ regions discovered by the Green Bank Telescope
\hii\ Region Discovery Survey (HRDS).  The HRDS sources are the most
distant yet discovered, and some nebulae are up to 20\,kpc from the
Sun.  Only 7\% of the HRDS nebulae are located at the near kinematic
distance and the average distance is 10.1\,kpc.  \hii\ regions
classified as ``bubbles'' have a similar distance distribution as
other classifications, in contrast to what previous authors have
assumed.

This work extends the spatial scale of previously known Galactic
structures.  The HRDS sources are concentrated at Galactocentric radii
of 4.25\,kpc and 6.0\,kpc, as is the sample of \hii\ regions known
prior to the HRDS.  When projected onto the Galactic plane, these
Galactocentric radius peaks appear as two concentric arc segments.  A
more complete discussion of the Galactic structure implications of the
present work is given in Bania et al. (2012, in prep.).

Kinematic distances are currently the only method for providing
distances to a large number of distant \hii\ regions.  Kinematic
distances are commonly thought to have large uncertainties.  Here we
assess the effect of three sources of uncertainty for kinematic
distances: differences in rotation curve models, non-circular motions,
and a change to the Solar circular rotation parameters.  We provide
quantitative maps of these uncertainties that will hopefully be of
great utility to future Galactic structure researchers.  The choice of
rotation curve and non-circular motions of magnitude 7\,\kms\ have a
similar effect on computed distances, while changing the Solar
circular rotation speed has a larger effect.  The combined
uncertainties are $\sim10\%$ for most of the Galactic zone studied
here ($67\degree > \ell > 18\degree$).

None of the basic Galactic structure results change as a result of
these uncertainties.  We analyzed the effect these uncertainties would
have on all known \hii\ regions in this zone of the Galaxy.  The
median absolute uncertainty is 0.5\,kpc, or $5\%$.  The median
difference between the near and the far distance is 6\,kpc for our
sample of \hii\ regions and therefore the resolution of the kinematic
distance ambiguity significantly improves our knowledge of the
Galactic location of a given \hii\ region.  We conclude that kinematic
distances are a reliable method for deriving distances over this zone
of the Galaxy.

\begin{acknowledgments}
Bob Rood, our friend and collaborator for many years, died on 2
November 2011.  The HRDS was partially supported by NSF award AST
0707853 to TMB.  \nraoblurb\ This research made use of NASA's
Astrophysics Data System Bibliographic Services.  Here we use
\hi\ data from the VLA Galactic Pane Survey (VGPS).  The VGPS is
supported by a grant from the Natural Sciences and Engineering
Research Council of Canada and from the National Science Foundation.
\nraoblurb
\end{acknowledgments}

\bibliographystyle{apj}
\bibliography{ref.bib}
\clearpage
\appendix
\section*{Appendix}
\setcounter{section}{0}
\section{The HRDS Web Site}
We have updated the HRDS website
described in Paper~II with results from the present work.  The site now
contains for each source the Figure~\ref{fig:spectra} \hiea\ spectra
and Figure~\ref{fig:channelmaps} single channel \hi\ images, as well
as data from Table~\ref{tab:data}.  We also provide an interactive
plot of the face-on map in Figure~\ref{fig:faceon}, and maps of the
total uncertainties in kinematic distances from
Figures~\ref{fig:big_nearfar} and \ref{fig:big_faceon}.  We will
continue to enhance this site as more is learned about the HRDS
sources.

\section{Distance Uncertainty Analysis\label{sec:appenb}}
We describe here our methodology for estimating kinematic
distance uncertainties associated
with the choice of rotation curve, streaming motions, and a change to
the Solar rotation speed.

\subsection{Uncertainties Caused by Choice of Rotation Curve}
There are many extant rotation curve models that one may choose when
deriving kinematic distances.  Three rotation curves commonly in use
today are those of B86, \citet[][hereafter C85]{clemens85}, and
\citet[][hereafter MGD07]{mcclure07}.  AB used the MGD07 curve for
their work.  All three curves assume that the distance from the
Sun to the Galactic center, $R_0$, is equal to 8.5\,kpc. 

All rotation curves have a Galactocentric range within which they are
applicable.  This range is set by the data that were used to create
the rotation curve.  C85 used CO data from the University of
Massachusetts-Stoney Brook survey \citep{sanders86}, \hi\ data from
\citet{burtongordon78}, and CO data measured in the direction of
\hii\ regions from \citet{blitz82}.  The data span $\sim 1-14$\,kpc.
The uncertainty of their model at the high end of this
range is large.  B86 used spectro-photometric distances of
\hii\ regions from \citet{brand88}, CO radial velocity measurements of
molecular clouds associated with these \hii\ regions from
\citet{brand87} and \citet{blitz82}, and \hi\ tangent point data from
\citet{fich89}.  They state that their curve is applicable within the
range $1.7$\,kpc to 17\,kpc.  MGD07 used
\hi\ tangent point data from the Southern Galactic Plane Survey
\citep[SGPS;][]{mcclure05}.  Their model is applicable over 3\,kpc --
8\,kpc.

We plot in the bottom panel of Figure~\ref{fig:rgal_theta} the
circular rotation speed versus Galactocentric distance for the B86
curve (solid line), the C85 curve (dashed line), and the MGD07 curve
(dotted line). In the top panel we show the standard deviation of the
three curves.  The shaded area shows the range over which the MGD07
curve is defined, 3\,kpc to 8\,kpc.  We extrapolate the MGD07 curve below
3\,kpc and assume a flat rotation curve above 8\,kpc.  By extending
this curve over the larger range of Galactocentric radii, we enable a
comparison between the three rotation curves over a larger portion of
the Galactic disk.  We will use these extrapolations for the analysis
below.  We caution however that the results in the Galactocentric
range over which we extrapolated should be viewed with some
skepticism.  Over 80\% of the HRDS nebulae with derived distances are
in the non-extrapolated region, as are 93\% of the AB sample.

Rotation curve models give kinematic distances for a given \lv\ pair
and we may therefore estimate the uncertainties associated with the
choice of a rotation curve for a grid of \lv\ locii.  We compute
for each rotation curve a grid of near distances and a grid of far
distances for a range of longitudes and velocities.  Each \lv\ grid
point therefore has a corresponding distance for the C85, B86, and
MGD07 rotation curves.  We consider longitudes in the range $80\arcdeg \ge
\ell \ge 10\arcdeg$ in increments of $0.1\arcdeg$ and velocities in
the range $200 \ge V_{\rm LSR} \ge -100\,\kms$ in increments of $0.1
\kms$.  We compute the standard deviation in the distances derived
with the three rotation curves for each \lv\ locus that is defined
in all three curves.  Finally, we calculate the percentage difference
from the B86 distance by dividing the standard deviation by the B86
distance.  We refer to this as the ``percentage uncertainty'' in a
distance based on the different rotation curves.

In Figure~\ref{fig:nearfar} we plot the percentage uncertainty in the
near (left panel) and the far (right panel) distances for our grid of
longitudes and velocities.  Each \lv\ locus in
Figure~\ref{fig:nearfar} has a corresponding uncertainty in both
panels.  For example, \lv$=(50\degree, 30\,\kms)$ has an uncertainty
of 15\% for the near distance and 4\% for the far distance; this
location is marked in Figure~\ref{fig:nearfar} with an ``x''.  
There
are two sets of curves shown in this figure.  For both sets, the
solid, dashed, and dotted curves represent the B86, C85, and MGD07
rotation curves, respectively.  One set of curves, running from
$\lv\simeq(10\degree, 150\,\kms)$ to $\lv\simeq(80\degree, 0\,\kms)$
shows the tangent point velocities for the three rotation curves.  The
other set of curves, spanning all longitudes near 0\,\kms, shows the
\lv\ locii where the near distance is zero for the three rotation
curves. The \lv\ locii enclosed in the gray area of
Figure~\ref{fig:nearfar} are defined in all three rotation curves.
The C85 and MGD07 curves are not defined for small LSR velocities at
high Galactic longitudes (they have distances $\leq 0$\,kpc).  This
effect causes the \lv\ area defined for all three curves to slant away
from $0\,\kms$ in the left panel of Figure~\ref{fig:nearfar}.

We find that the percentage uncertainties are generally greater for
near distances than for far distances.  
Uncertainties are especially
large, $>20\%$, near the Sun (at low LSR velocities) and at higher
longitudes.  Uncertainties in the far distance are mostly $<\,5\%$,
but increase to about $\sim\,10\%$ at higher longitudes.  

We plot the rotation curve uncertainties from Figure~\ref{fig:nearfar}
projected onto the Galactic plane in Figure~\ref{fig:faceon_three}.
To construct this face-on map, we find for each \lv\ locus the
corresponding distance using the B86 curve.  We then use the
corresponding percentage uncertainty from Figure~\ref{fig:nearfar} at
each \lv\ locus for the value in the face-on map.  In Figure
\ref{fig:faceon_three}, the tangent point location is the solid black
line and the Solar orbit is indicated with a dashed light gray line.
We plot the longitude range of the HRDS with solid white lines.  The
white holes in Figure~\ref{fig:faceon_three} correspond to
\lv\ locii in the B86 curve that are not defined by the other two
rotation curves (see below).

Not all \lv\ locii are defined for all three rotation curves.  The
regions undefined in other curves that are defined in the B86 curve
are identifiable as the \lv\ locii in Figure~\ref{fig:nearfar} in
between the gray filled region and the B86 tangent point velocity
curve.  For example, the tangent point velocity for the C85 curve is
significantly less than that of the B86 and MGD07 curves near $l =
20\arcdeg$.  This leads to one of the undefined white holes in
Figure~\ref{fig:faceon_three}.

With the exception of locations within $\sim 1\,$kpc of the Sun and
Galactic longitude $\gtrsim\,50\arcdeg$, the choice of rotation curve
is not a significant source of uncertainty when computing kinematic
distances.  Distance variations associated with the choice of rotation
curve are generally small; for $\sim70\%$ of the defined Galactic
locations considered, the differences in distance are $<5\%$.  Over
94\% of the defined locations have distance differences $\leq 10\%$,
and over 99\% of the defined locations have distance differences $\leq
20\%$.


\subsection{Uncertainties Caused by Non-Circular Motions}
There are two main sources of ``non-circular motions'': systematic
velocity fields within a source and ordered large-scale Galactic
streaming motions.  The Galactic Bar and the 3\,Kpc Arm for example
produce streaming motions that occur throughout the inner Galaxy.  We
estimate the uncertainties caused by non-circular motions by
recomputing the distances found using the B86 curve using our grid of
longitudes, but adding 7\,\kms\ and subtracting 7\,\kms\ to the
velocity grid.  We use 7\,\kms\ as an estimate of the true streaming
motions, which may be 5 to 10\,\kms\ \citep{burton66} and do not
include any estimate of the contribution from systematic flows within the
source.

Streaming motions are of course not random, as we have assumed here.
They are associated with large-scale Galactic features and therefore
are present for distinct areas of \lv-space.  Our estimates give
order-of-magnitude values for the effect of streaming motions.  They
do not, however, provide error estimates for any specific nebula in
our sample.

We compute for each \lv\ locus three grids of kinematic distances
using the B86 curve: one grid with no velocity offset, one grid where
each locus is shifted by $+7$\,\kms, and one grid where each
locus is shifted by $-7$\,\kms.  We then compute the percentage
uncertainty for each \lv\ locus as before by dividing the standard
deviation at each \lv\ grid locus by the B86 distance.  As before,
we project the \lv\ percentage uncertainties onto the Galactic plane.

We plot in Figure~\ref{fig:nearfar_stream} the uncertainties from
random streaming motions of magnitude 7\,\kms.  The shaded areas and
curves here have the same meaning as in Figure \ref{fig:nearfar}, but
we only plot the curves for the B86 rotation curve.  Locations within
7\,\kms\ of the tangent point velocity are undefined since adding
$7\,\kms$ results in a velocity greater than the tangent point
velocity.  As before, the uncertainties in the near distances are
greater than those of the far distances, and uncertainties are greater
near the Sun.

We transform the data of Figure~\ref{fig:nearfar_stream} as
before into the face-on plot of Figure~\ref{fig:faceon_stream}.  The
lines and curves in Figure~\ref{fig:faceon_stream} are as in
Figure~\ref{fig:faceon_three}.  The zone corresponding to velocities
within 7\,\kms\ of the tangent point velocity is undefined and we
therefore leave it blank.  Although there are \lv\ locii near
0\,\kms\ that are similarly undefined for near distances in
Figure~\ref{fig:nearfar_stream}, these locii are defined for the
far distances and therefore there are no holes near the Solar orbit in
Figure~\ref{fig:faceon_stream}.

With the exception of distances within a few kpc of the Sun, randomly
distributed $\pm$\,7\,km\,s$^{-1}$ non-circular motions are not a
significant source of uncertainty when computing kinematic distances
over the longitude range studied here.  For $\sim85\%$ of the defined
Galactic locations the distance uncertainties are $<10\%$.  Over 95\%
of the defined locations have distance uncertainties $\leq 20\%$.
Distance uncertainties associated with non-circular motions of
7\,\kms\ are generally $5-10\%$.  Both in magnitude, and in the
\lv\ locii, the uncertainties due to non-circular motions are
similar to those associated with the choice of rotation curve.

\subsection{Uncertainties Caused by a Change of Solar Rotation Parameters}
Finally, we estimate the effect on the derived kinematic distances of
a change in the IAU standard value for the Solar circular rotation
speed, $\Theta_0$.  \citet{reid09} recommended revised values for the
distance from the Sun to the Galactic center, $R_0$, and for
$\Theta_0$ based on their observations of the parallax of Galactic
masers associated with massive star formation.  Their observations
support a distance from the Sun to the Galactic center of
$8.4\pm0.6$\,kpc, and a Solar circular rotation speed of
$254\pm16\,\kms$.  Since their value for the distance to the Galactic
center is consistent with the IAU standard value of 8.5\,kpc, we do
not include this change in the following analysis.

We compute for each \lv\ locus two grids of kinematic distances
using the B86 rotation curve: one with $\Theta_0=220\,\kms$ and one
with $\Theta_0=250\,\kms$.  We then compute the percentage uncertainty
for each \lv\ locus as before by dividing the standard deviation of
these two \lv\ grids by the distance computed with
$\Theta_0=220\,\kms$; we project these percentage uncertainty grids
onto the Galactic plane.



\begin{figure}
\centering
\includegraphics[width=2.9 in]{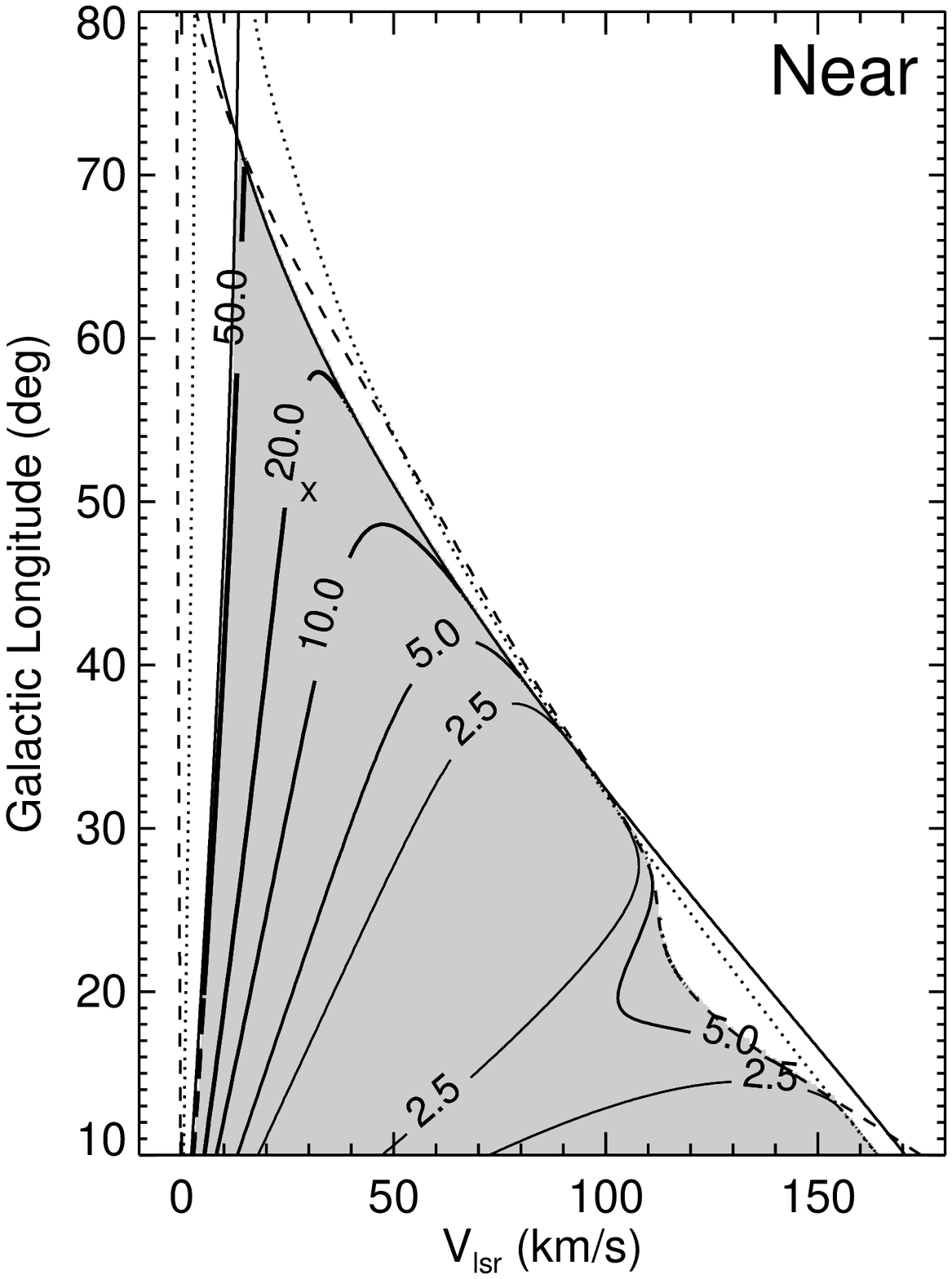}
\includegraphics[width=2.9 in]{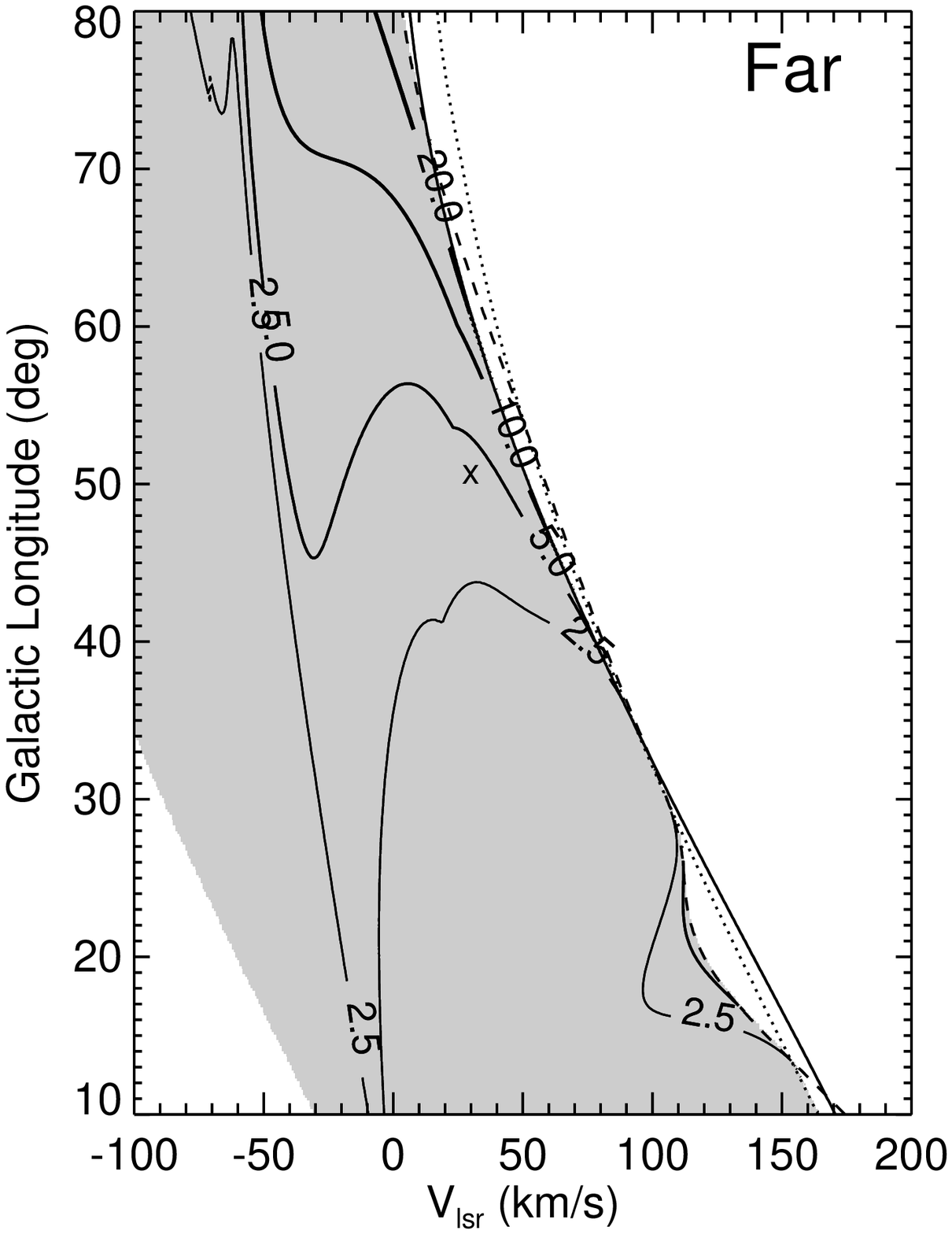}

\caption{The percentage uncertainty in kinematic distances associated
  with the choice of rotation curve.  Each \lv\ locus gives the
  standard deviation in the distances computed using the C85, B86, and
  MGD07 rotation curves divided by the distance calculated using the
  B86 rotation curve.  We mark the example position mentioned in the
  text, \lv$=(50\degree, 30\,\kms)$, with an ``x''.  The left panel shows the
  percentage uncertainty in near distances and the right panel shows
  the same for far distances.  The gray area contains \lv\ locii
  that are defined for all three rotation curves whereas white areas
  are undefined for one or more rotation curves.  The three curved
  lines (not including the contours) show the \lv\ locii where the
  near distance is zero, and also the \lv\ locii of the tangent
  point for the B86 curve (solid), C85 curve (dashed), and the MGD07
  curve (dotted).}

\label{fig:nearfar}
\end{figure}

We plot in Figure~\ref{fig:nearfar_reid} the percentage uncertainty in
the near (left panel) and far distances (right panel).  The shaded
areas have the same meaning as in Figure~\ref{fig:nearfar}.  The
curves in Figure~\ref{fig:nearfar_reid} show the tangent point
velocities and velocities at which the near distance is zero, as
before.  The solid line plots the B86 curve with $\Theta_0 =220\,\kms$
and the dotted line plots the effect on the tangent point velocities
and velocities at which the near distance is zero when $\Theta_0$ is
changed to 250\,\kms.

\begin{figure}
\centering
\includegraphics[width=3.5 in]{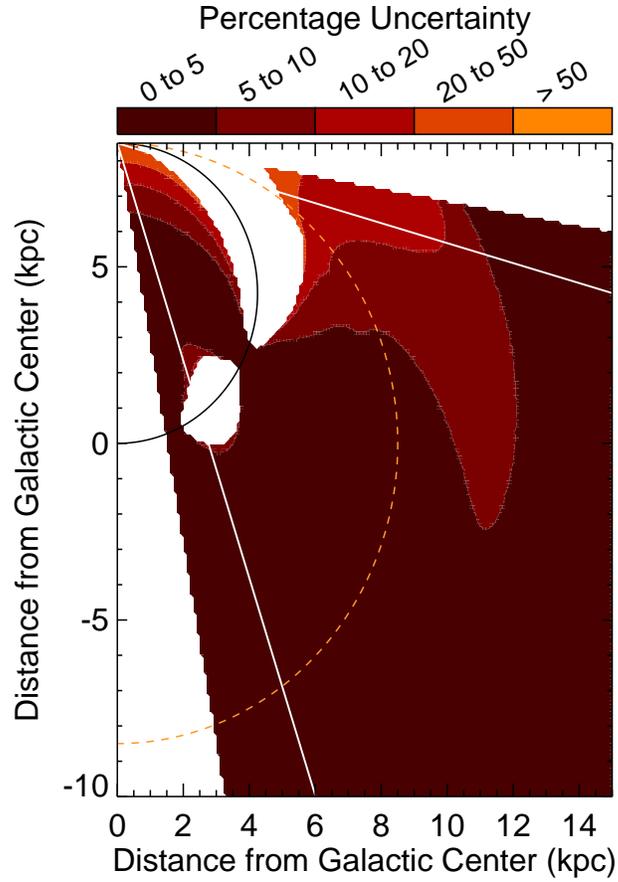}

\caption{Face-on map of the percentage uncertainty in the three
  rotation curve models.  The tangent point distance is shown as the
  solid curve and the Solar orbit is shown as the dashed curve.  The
  range of the HRDS sources with assigned distances is shown with
  straight lines.  The white holes in the figure shows areas that are
  not defined for all three rotation curves.}

\label{fig:faceon_three}
\end{figure}

\begin{figure}
\centering
\includegraphics[width=2.9 in]{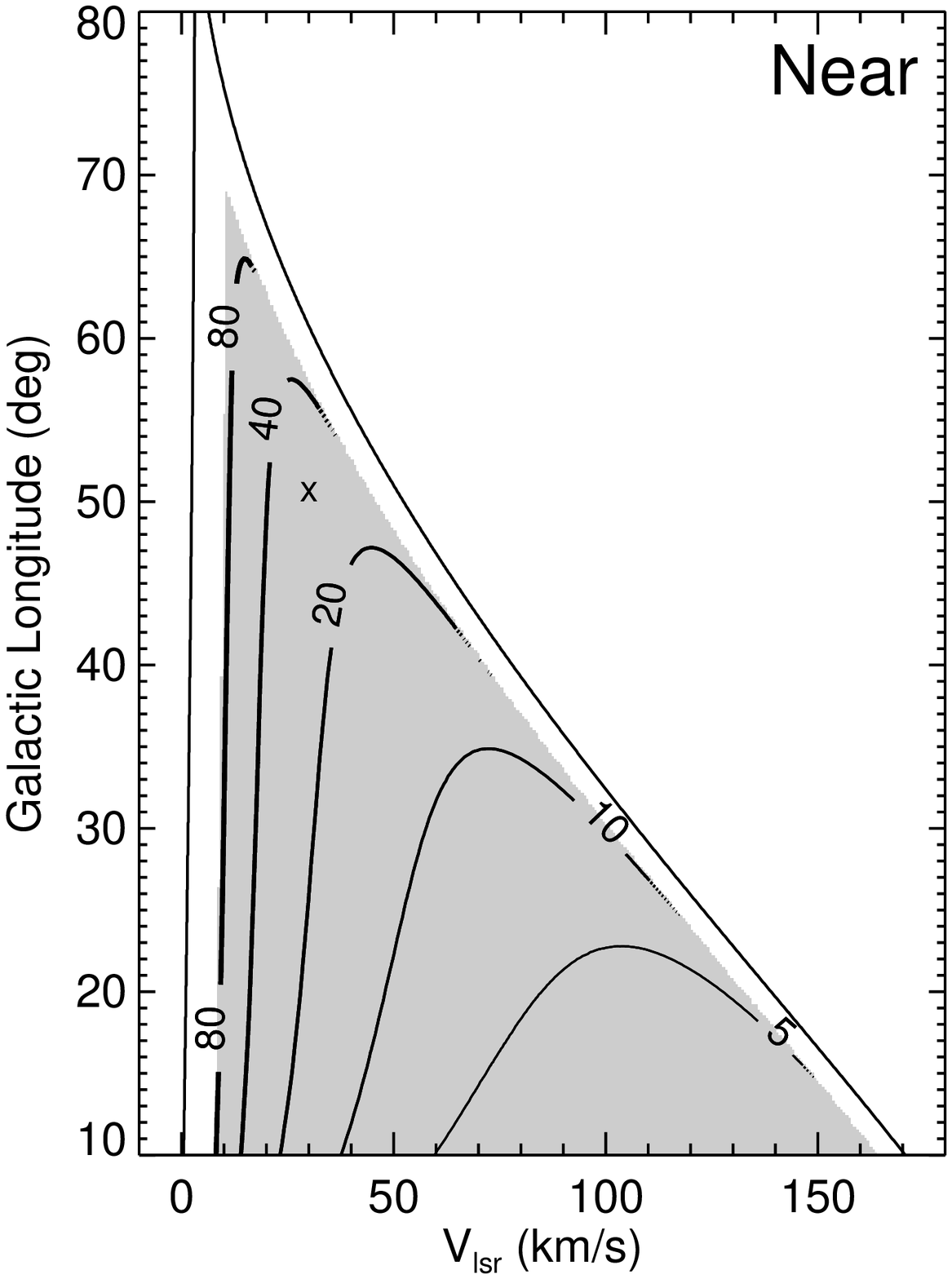}
\includegraphics[width=2.9 in]{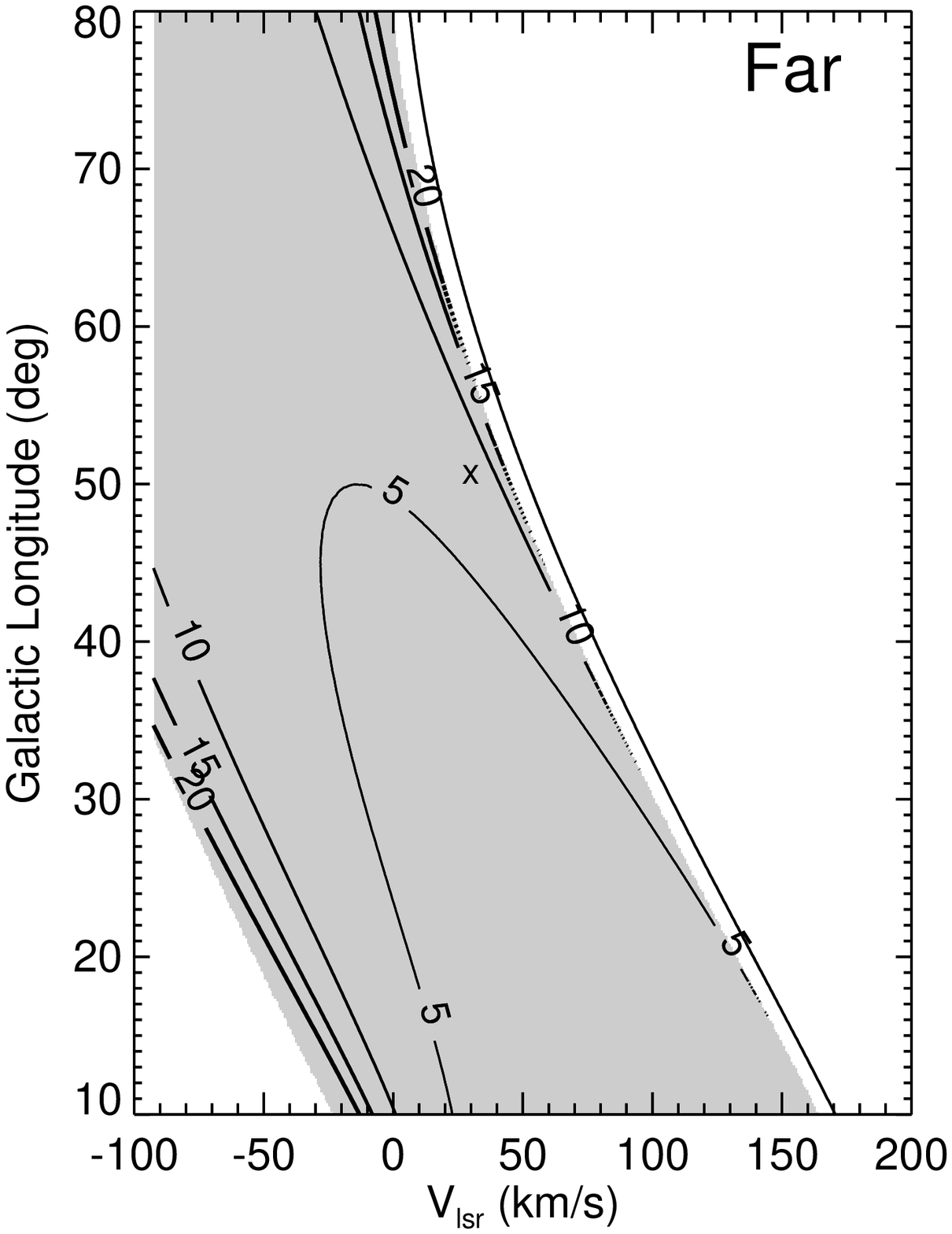}

\caption{The percentage uncertainty in kinematic distances associated
  with non-circular motions of $7\,\kms$ for near (left panel) and far
  distances (right panel).  The solid black lines show the
  \lv\ locii where the near distance is zero, and also the
  \lv\ locii of the tangent point for the B86 curve.  We mark the
  example position mentioned in the text, \lv$=(50\degree, 30\,\kms)$, with an
  ``x''.}

\label{fig:nearfar_stream}
\end{figure}

\begin{figure}
\centering
\includegraphics[width=3.5 in]{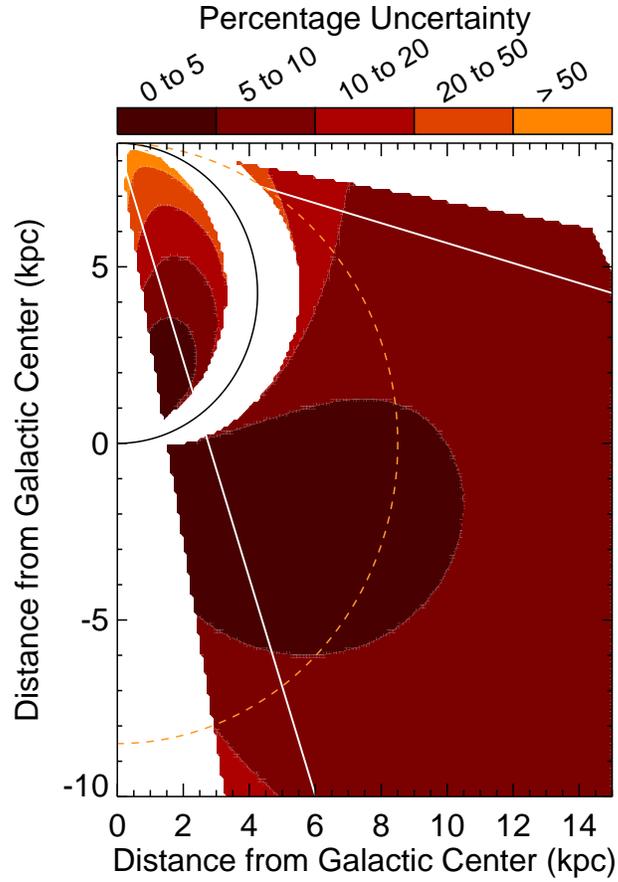}

\caption{Face-on map of the percentage uncertainty in kinematic
  distances associated with non-circular motions of 7\,\kms.  The meaning
  of the curves and lines are the same as in
  Figure~\ref{fig:faceon_three}.  No comparison may be made within
  7\,\kms\ of the tangent point velocity, which leads to the white area
  surrounding the tangent point distance where no uncertainties are
  calculated.}

\label{fig:faceon_stream}
\end{figure}

\begin{figure}
\centering
\includegraphics[width=2.9 in]{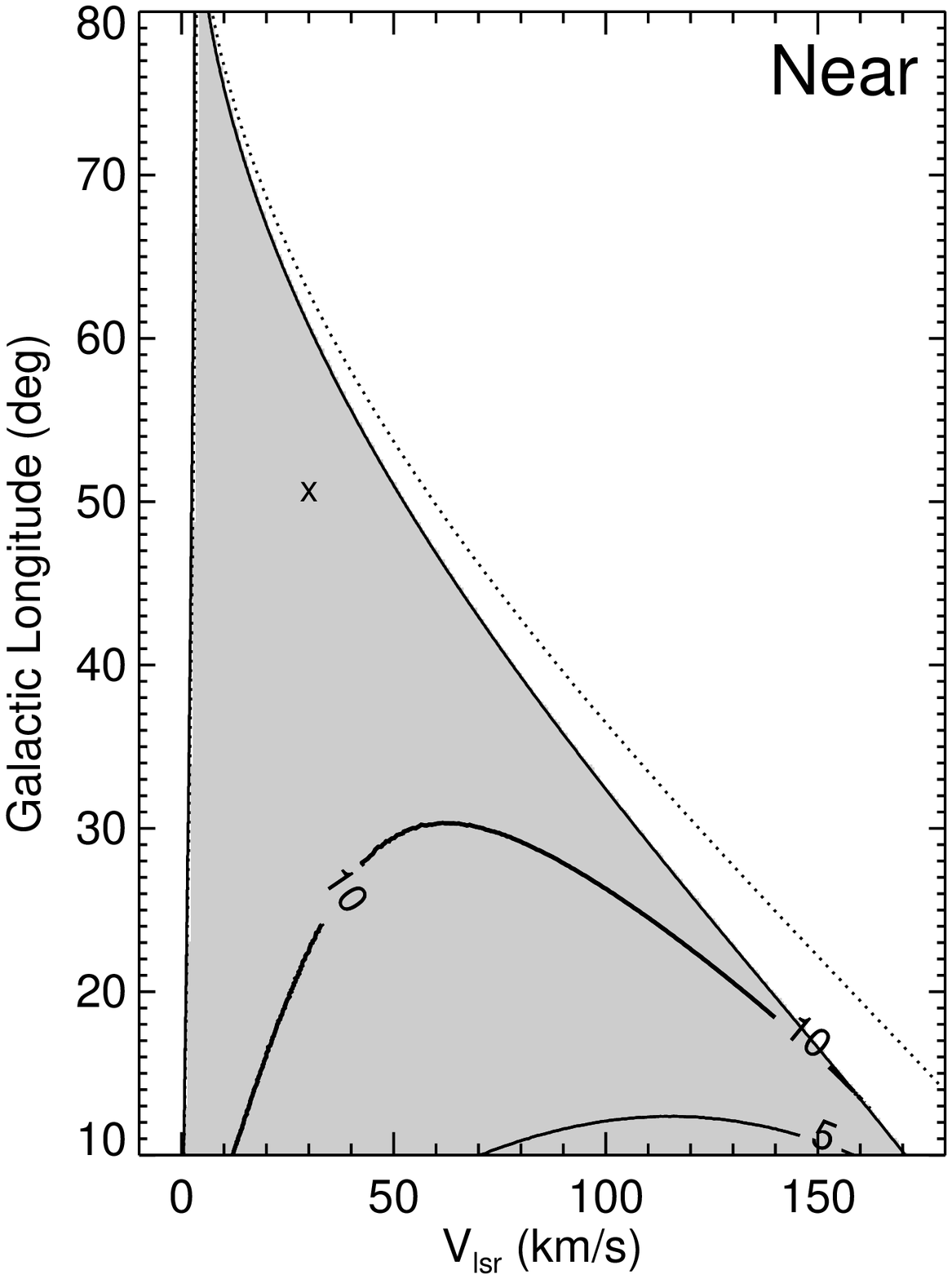}
\includegraphics[width=2.9 in]{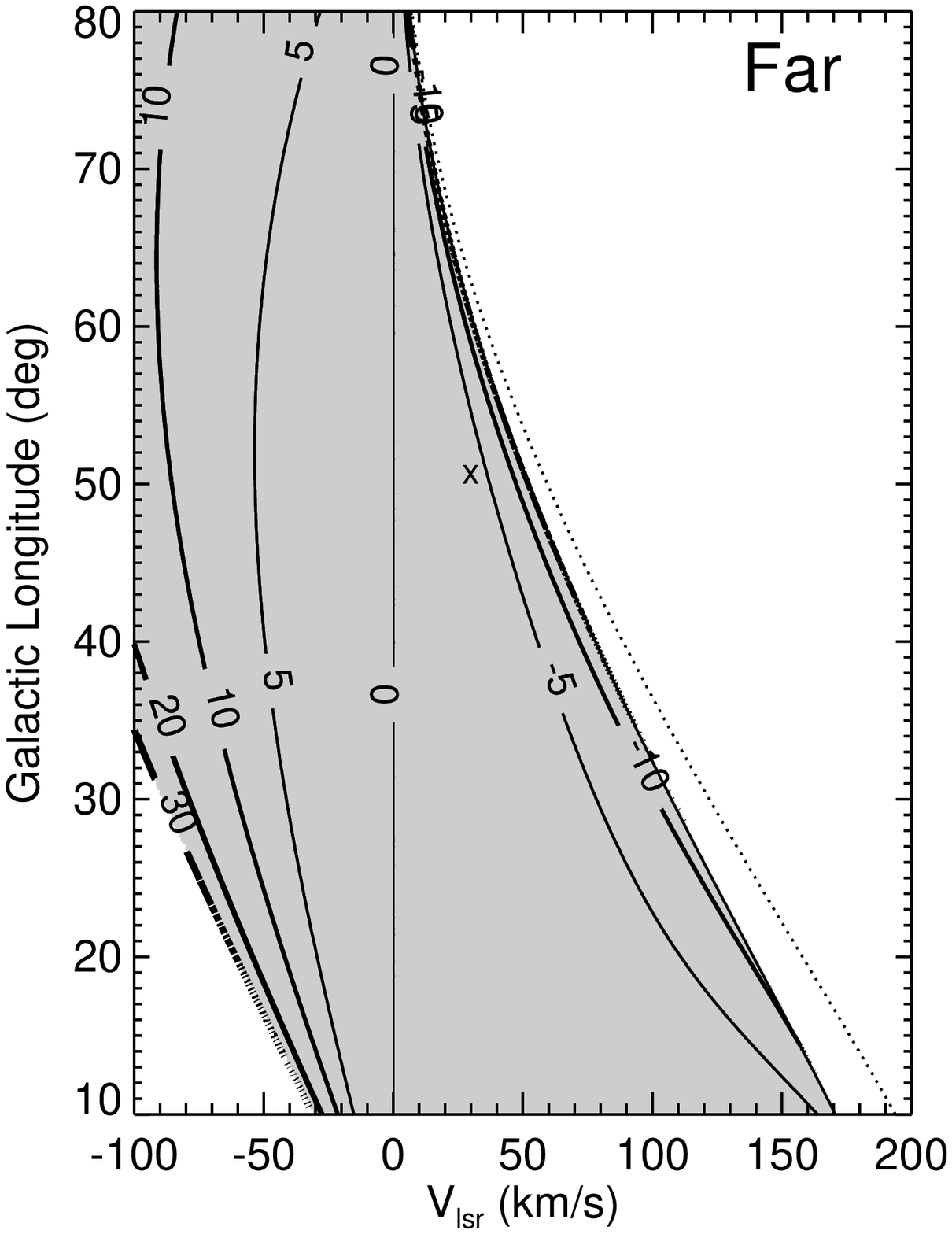}

\caption{The percentage uncertainty associated with changing the
  circular Solar rotation speed from 220\,\kms\ to 250\,\kms\ for near
  (left panel) and far distances (right panel).  The black lines show
  the \lv\ locii where the near distance is zero, and also the
  \lv\ locii of the tangent point for the B86 curve with $\Theta_0
  = 220\,\kms$.  The dotted line shows the same for $\Theta_0 =
  250\,\kms$.  We mark the example position mentioned in the text,
  \lv$=(50\degree, 30\,\kms)$, with an ``x''.}

\label{fig:nearfar_reid}
\end{figure}

\begin{figure}
\centering
\includegraphics[width=3.5 in]{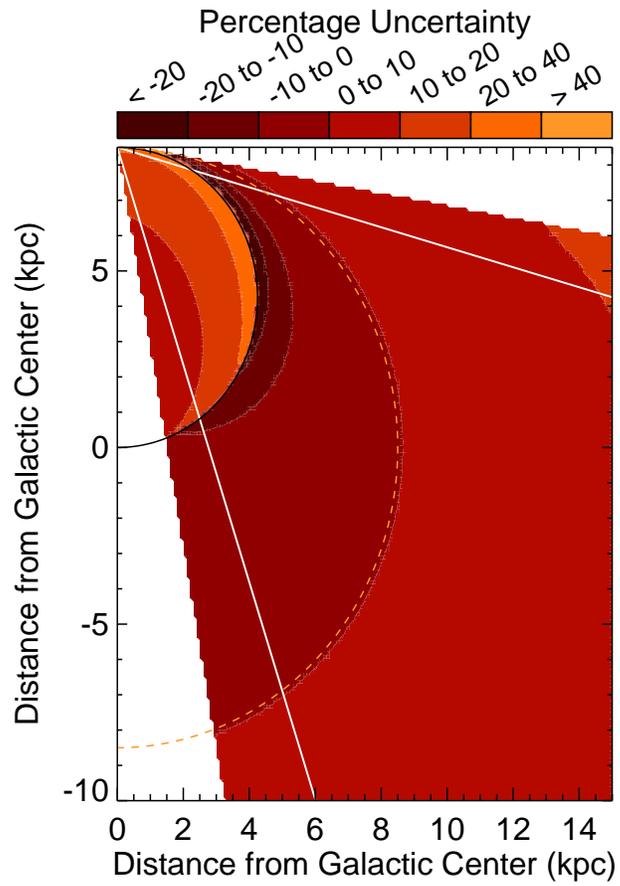}

\caption{Face-on map of the percentage uncertainty caused by changing
  the Solar rotation speed to 250\,\kms.  The meaning of the curves
  and lines are as in Figure~\ref{fig:faceon_three}.}

\label{fig:faceon_reid}
\end{figure}

We transform the data from in Figure~\ref{fig:nearfar_reid} as
before into the face-on plot of Figure~\ref{fig:faceon_reid}.  The
lines and curves in Figure~\ref{fig:faceon_reid} are as in
Figure~\ref{fig:faceon_three}.  There are no undefined areas of this
figure because all \lb\ locii defined with
$\Theta_0=220$\,\kms\ are defined with $\Theta_0$=250\,\kms\ (the
inverse is not true though).

In general, the uncertainties associated with changing the Solar
rotation speed are greater than the uncertainties associated with
either the selection of a rotation curve or with non-circular motions.
Changing the IAU standard for $\Theta_0$ results in differences $\leq
10\%$ for most \lv\ locii.  Changing the Solar rotation speed
results in distance uncertainties of up to 10\% for $\sim75\%$ of the
defined Galactic locations.  Almost 95\% of the defined Galactic
locations have distance uncertainties $\leq 20\%$.  One of the main
effects of changing the Solar rotation speed is that the tangent point
velocity increases in the first Galactic quadrant.  This leads to the
large uncertainties near the tangent point distance.  Changing the
Solar circular rotation speed causes distance uncertainties of $\sim
10\%$, which is generally greater than the uncertainties associated
with the choice of rotation curve and the effect of non-circular
motions.

\begin{deluxetable}{lcccccccccccccc}
\tabletypesize{\scriptsize}
\tablecaption{HRDS Kinematic Distances}
\tablewidth{0pt}
\tablehead{

\colhead{Source} & 
\colhead{$\ell$} &
\colhead{$b$} &
\colhead{$V_{\rm lsr}$} &
\colhead{$V_{\rm max}$} & 
\colhead{$V_{\rm TP}$} & 
\colhead{D$_{\rm N}$} &
\colhead{D$_{\rm F}$} &
\colhead{N/F} & 
\colhead{QF} & 
\colhead{$D_\sun$} &
\colhead{$\sigma_D$} &
\colhead{R$_{\rm gal}$} &
\colhead{$z$} \\

\colhead{} &
\colhead{deg.} &
\colhead{deg.} &
\colhead{\kms} &
\colhead{\kms} &
\colhead{\kms} &
\colhead{kpc} &
\colhead{kpc} &
\colhead{} &
\colhead{} &
\colhead{kpc} &
\colhead{kpc} &
\colhead{kpc} &
\colhead{pc}
}

\startdata
G017.928$-$0.677  & $17.928$ & $-0.677$             & \phantom{$-$}$39.1$ & \phantom{0}53 & $145.7$ & $3.4$   & $12.8$ & F       & B & 12.8           & 0.5     & 5.4 & $-$150                     \\
G018.077+0.071    & $18.077$ & \phantom{$-$}$0.071$ & \phantom{$-$}$58.2$ & 128           & $145.2$ & $4.4$   & $11.8$ & F       & B & 11.8           & 0.4     & 4.5 & \phantom{$-$}\phantom{0}15 \\
G018.097$-$0.324  & $18.097$ & $-0.324$             & \phantom{$-$}$50.8$ & \nodata       & $145.1$ & $4.0$   & $12.1$ & \nodata & C & \nodata        & \nodata & 4.8 & \nodata                    \\
G018.156+0.099    & $18.156$ & \phantom{$-$}$0.099$ & \phantom{$-$}$53.0$ & \phantom{0}52 & $145.0$ & $4.1$   & $12.0$ & N       & A & \phantom{0}4.1 & 0.4     & 4.8 & \phantom{$-$}\phantom{0}71 \\
G018.236+0.395    & $18.236$ & \phantom{$-$}$0.395$ & \phantom{0}$-0.4$   & \phantom{0}51 & $144.7$ & \nodata & $16.3$ & F       & A & 16.3           & 1.4     & 8.7 & \phantom{$-$}110           \\
G018.324+0.026    & $18.324$ & \phantom{$-$}$0.026$ & \phantom{$-$}$50.4$ & 125           & $144.4$ & $4.0$   & $12.2$ & F       & A & 12.2           & 0.4     & 4.9 & \phantom{$-$}\phantom{0}55 \\
G018.584+0.344    & $18.584$ & \phantom{$-$}$0.344$ & \phantom{$-$}$10.8$ & 112           & $143.6$ & $1.1$   & $15.0$ & F       & B & 15.0           & 0.9     & 7.4 & \phantom{$-$}\phantom{0}90 \\
G018.630+0.309    & $18.630$ & \phantom{$-$}$0.309$ & \phantom{$-$}$14.0$ & \nodata       & $143.4$ & $1.5$   & $14.7$ & \nodata & C & \nodata        & \nodata & 7.1 & \nodata                    \\
G018.708$-$0.126  & $18.708$ & $-0.126$             & \phantom{$-$}$60.5$ & \nodata       & $143.2$ & $4.4$   & $11.7$ & \nodata & C & \nodata        & \nodata & 4.5 & \nodata                    \\
G018.751+0.254    & $18.751$ & \phantom{$-$}$0.254$ & \phantom{$-$}$19.1$ & 125           & $143.0$ & $1.9$   & $14.2$ & F       & A & 14.2           & 0.7     & 6.7 & \phantom{$-$}\phantom{0}63 \\

\enddata

\label{tab:data}
\tablecomments{Table~\ref{tab:data} is published in its entirety in the
electronic edition of the {\it Astrophysical Journal}.  A portion is
shown here for guidance regarding its form and content.}
\end{deluxetable}
\clearpage

\clearpage

\clearpage


\end{document}